\documentclass[12pt,a4paper, amsmath,amssymb]{article}
\usepackage[margin=0.8in]{geometry}
\usepackage{amssymb}
\usepackage{amsmath}
\usepackage{amsfonts}
\usepackage{bm}
\usepackage[utf8]{inputenc}
\usepackage{graphicx}
\usepackage{float}
\usepackage[colorlinks,linkcolor = blue,
urlcolor  = blue,
citecolor = blue,
anchorcolor = blue]{hyperref}
\usepackage{xcolor}
\usepackage{makecell}
\usepackage{ctable} 
\usepackage{adjustbox}
\def\thefootnote{\fnsymbol{footnote}}
\usepackage[colorlinks,linkcolor = blue,
urlcolor  = blue,
citecolor = blue,
anchorcolor = blue]{hyperref}
\usepackage{cite}
\begin{document}
\begin{center}
{\Large \textbf{Explaining the 96 GeV Di-photon Anomaly\\[0.25cm]
 in a Generic 2HDM Type-III 
}}
\thispagestyle{empty}
%
\vspace{1cm}

{\sc
R. Benbrik$^1$\footnote{\url{r.benbrik@uca.ac.ma}},
M. Boukidi$^1$\footnote{\url{mohammed.boukidi@ced.uca.ma}},
S. Moretti$^{2,3}$\footnote{\url{S.Moretti@soton.ac.uk}; \url{stefano.moretti@physics.uu.se}},
S. Semlali$^2$\footnote{\url{souad.semlali@soton.ac.uk}}\\
}
\vspace{1cm}
{\sl
  $^1$Polydisciplinary Faculty, Laboratory of Fundamental and Applied Physics, Cadi Ayyad University, Sidi Bouzid, B.P. 4162, Safi, Morocco.\\
\vspace{0.1cm}
 $^2$School of Physics and Astronomy, University of Southampton, Southampton, SO17 1BJ, United Kingdom.\\
\vspace{0.1cm}
$^3$Department of Physics and Astronomy, Uppsala University, Box 516, SE-751 20 Uppsala, Sweden.
}
\end{center}
\vspace*{0.1cm}
\begin{abstract}
  Motivated by results recently reported by the CMS Collaboration about an excess in the di-photon spectrum at about 96 GeV, especially when combined with another long-standing  anomaly at the same value in the $b\bar b$ invariant mass spectrum in four-jet events collected at LEP, we show that a possible explanation to both phenomena can be found at 1$\sigma$ level in a generic 2-Higgs Doublet Model (2HDM) of Type-III in presence of a specific Yukawa texture, wherein Lepton Flavour Violating (LFV)  (neutral) currents are induced at  tree level. Bounds from Higgs data play a major role in limiting the parameter space of this scenario, yet
we find solutions with  $m_H = 125$ GeV and $m_h = 96$ GeV consistent with current theoretical and experimental bounds.
\end{abstract}

\def\thefootnote{\arabic{footnote}}
\setcounter{page}{0}
\setcounter{footnote}{0}

\newpage

\section{Introduction}
The discovery of a Higgs boson, $H$,  compatible with the one predicted by the Standard Model (SM), carrying a mass of 125 GeV, at the Large Hadron Collider (LHC) \cite{Aad:2012tfa,Chatrchyan:2012ufa}, has widely been regarded as the opening of a new age in particle physics. Such a Higgs boson is the first scalar and fundamental particle to be found in Nature and the last remaining undiscovered particle required for the experimental confirmation of the SM. Even though the measurements of the production and decay rates of the 125 GeV Higgs boson are currently in good agreement with the expectations within the SM, the Higgs sector chosen by Nature may not necessarily be the minimal one of such a construct. There might be additional real or complex singlets, doublets triplets, and so on, or any mixture of these. An extended Higgs sector may also be incorporated into a proper theoretical framework, for example, a specific realisation of the 2-Higgs Doublet Model (2HDM) can be made part of the Minimal Supersymmetric Standard Model (MSSM) \cite{Gunion:1992hs,Branco:2011iw}. 

Di-photon signatures originating from the production and decay of new (pseudo)scalar Higgs states are of special importance  since this was the channel used to claim the aforementioned discovery. In Ref. \cite{CMS:2018cyk}, CMS has identified two excesses in the search for light neutral states in $pp\to \phi \to \gamma\gamma$  with 2$\sigma$ local significance at $m_{\gamma\gamma} = 97.6$ GeV at Run-1 (2012)  with 19.7 fb$^{-1}$ of luminosity and with 3$\sigma$ local significance at $m_{\gamma\gamma} = 95.3$ GeV at Run-2 (2016) with 35.9 fb$^{-1}$ of luminosity.  Given the ${\cal O}(1$ GeV) resolution of the CMS detector in di-photon 
final states, clearly, these two excesses are consistent with the same mass point. Furthermore, 
while not confirmed or denied by the ATLAS Collaboration yet, most recently, CMS has reported an additional local excess with a significance of 3.1 $\sigma$ (for $m_h = 100 $ GeV)  \cite{CMS:2022rbd} in the di-tau final state in the search for a light Higgs boson. If this measurement is interpreted instead as further evidence of a 96 GeV resonance,   it corresponds to
a 2.6$\sigma$ deviation. 

There seems to be therefore a compelling case to address the question as to whether any theoretical model can explain these 
 measurements. Indeed, this have already resulted in several interesting academic papers, which explored some realisations that could simultaneously explain not only the above $\gamma\gamma $ excesses, while being consistent with the current measurements related to the observed Higgs boson in 2012
\cite{Cao:2016uwt,Heinemeyer:2021msz,Biekotter:2019kde,Biekotter:2021qbc,Biekotter:2022jyr,Cacciapaglia:2016tlr,Cline:2019okt,Cao:2019ofo,Biekotter:2021ovi,Crivellin:2017upt,Biekotter:2022abc,Abdelalim:2020xfk,ALEPH:2006tnd}, but also an anomaly which has remained from LEP data, in the $e^+e^-\to Z(H\to b\bar{b})$ channel, wherein an excess was seen in the $b\bar{b}$ invariant mass, again around 98 GeV \cite{LEP-Excess}. Indeed, given the resolution of di-jet masses at LEP, this number is also consistent with the location of the CMS excesses. 

Here, we intend to study such reported excesses in the context of a 2HDM Type-III with a specific Yukawa texture allowing for
some amount of  Lepton Flavour Violation (LFV) \cite{Crivellin:2013wna}.  This is done in order to comply with LEP, Tevatron, LHC and low energy measurements sensitive to not only the SM-like Higgs boson $H$ but also to a potentially lighter $h$ state playing the role of
the object behind the aforementioned excesses at 96 GeV. In fact,  a generic 2HDM contains five physical (pseudo)scalar particles, including two CP-even states ($h$ and $H$, with $m_h<m_H$), one CP-odd one ($A$) and a charged Higgs boson pair ($H^\pm$). Based on the way which the Higgs doublets couple to the fermions, the 2HDM is categorised into Type-I, -II, lepton-specific  and flipped. The 2HDM Type-III corresponds to the case where each of the two Higgs doublets couples to all fermions simultaneously. As a consequence, tree-level Flavour Changing Neutral Currents (FCNCs) in the sectors of charged quarks and leptons are being induced. In our approach, rather than postulating a $Z_2$ symmetry (exact or softly-broken) to control the latter, we assume a 
generic Yukawa texture  that we will constrain by exploiting theoretical conditions  of self-consistency as well as experimental measurements  of masses and couplings.   

The plan of our paper is as follows. In the next section, we describe the theoretical framework. We then move on to explain the features of the two experimental excesses (LEP and LHC). Then we will map one onto the others by presenting numerical results. We will finally conclude.

\section{Generic 2HDM Framework }

The 2HDM is one of the simplest extension of the SM as it  consists of two complex doublets of Higgs  fields $\Phi_i$ ($i =1, 2$) with hypercharge $Y = +1$. The  $SU(2)_L$$\otimes$$ U(1)_Y$ invariant scalar potential is given by \cite{Branco:2011iw}:
\begin{eqnarray}
  \mathcal{V} &= m_{11}^2 \Phi_1^\dagger \Phi_1+ m_{22}^2\Phi_2^\dagger\Phi_2 - \left[m_{12}^2
	\Phi_1^\dagger \Phi_2 + \rm{H.c.}\right] + \lambda_1(\Phi_1^\dagger\Phi_1)^2 +
	\lambda_2(\Phi_2^\dagger\Phi_2)^2 +
	\lambda_3(\Phi_1^\dagger\Phi_1)(\Phi_2^\dagger\Phi_2)  ~\nonumber\\ &+
	\lambda_4(\Phi_1^\dagger\Phi_2)(\Phi_2^\dagger\Phi_1) +
	\frac12\left[\lambda_5(\Phi_1^\dagger\Phi_2)^2 +\rm{H.c.}\right] 
 +\left\{\left[\lambda_6(\Phi_1^\dagger\Phi_1)+\lambda_7(\Phi_2^\dagger\Phi_2)\right]
	(\Phi_1^\dagger\Phi_2)+\rm{H.c.}\right\} \label{C2HDMpot}
\end{eqnarray}
Following the hermiticity of the potential given by eq. (\ref{C2HDMpot}), $m_{11}^2$, $m_{22}^2$ and $\lambda_{1,2,3,4}$ are real parameters whereas $\lambda_{5,6,7}$ and $m_{12}^2$ can be complex. Adopting the CP-conserving 2HDM, $\lambda_{5,6,7}$ and $m_{12}^2$ are real. Keeping $\lambda_{6,7}$ terms in the potential, there are direct contribution of $\lambda_{6,7}$ to the $ h\gamma\gamma$ process via the triple Higgs coupling $h H^+ H^-$. However, in this study, the contribution of charged Higgs bosons to such a vertex is negligible so that the effect of $\lambda_{6,7}$ is very small. We then set $\lambda_{6,7} = 0$ for the whole analysis. Under the above conditions, the potential has seven  independent parameters: $m_h, m_H, m_A, m_{H^\pm},\tan\beta (=v_2/v_1), \sin(\beta - \alpha)$ and $m^2_{12}$. In this work we consider $H$ as the observed SM-like boson with $m_H = 125$ GeV, so that only six free parameters are left. Here, $v_1$ and $v_2$ are the Vacuum Expectation Values (VEVs) of the two Higgs doublets while $\alpha$ is the mixing angle in the CP-even Higgs sector
 \cite{Branco:2011iw}.  

As for the Yukawa sector, the general scalar to fermions couplings are expressed by:
\begin{eqnarray}
-{\cal L}_Y &=& \bar Q_L Y^u_1 U_R \tilde \Phi_1 + \bar Q_L Y^{u}_2 U_R
\tilde \Phi_2  + \bar Q_L Y^d_1 D_R \Phi_1 
+ \bar Q_L Y^{d}_2 D_R \Phi_2 \nonumber \\
&+&  \bar L Y^\ell_1 \ell_R \Phi_1 + \bar L Y^{\ell}_2 \ell_R \Phi_2 + H.c. 
\label{eq:Yu}
\end{eqnarray}

Before the Electro-Weak Symmetry Breaking (EWSB), all $Y^{f}_{1,2}$ are arbitrary $3\times 3$ matrices and fermions are not physical eigenstates. Therefore, we have the freedom to choose $Y^u_1$, $Y^d_2$, and $Y^\ell_2$ to have diagonal forms; that is, $Y^u_1={\rm diag}(y^u_1, y^u_2, y^u_3)$ and 
$Y^{d,\ell}_{2}= {\rm diag}(y^{d,\ell}_1, y^{d,\ell}_2, y^{d\ell}_{3})$. In this study, we investigate 2HDM Type-III, where neither a global symmetry is imposed on the Yukawa sector nor an alignment in the flavour space is enforced. We adopt instead the Cheng-Sher ansatz~\cite{Cheng:1987rs, Diaz-Cruz:2004wsi}, which assumes a flavour symmetry that suggests a specific texture of the Yukawa matrices, where FCNC effects are proportional to the masses of the fermions and dimensionless real parameters, with $\tilde{Y}_{ij} \propto \sqrt{m_i m_j}/ v ~\chi_{ij}$. After EWSB, the Yukawa Lagrangian can be written, in terms of the mass eigenstates of the Higgs bosons, as follows:
\begin{align}
	-{\cal L}^{III}_Y  &= \sum_{f=u,d,\ell} \frac{m^f_j }{v} \times\left( (\xi^f_h)_{ij}  \bar f_{Li}  f_{Rj}  h + (\xi^f_H)_{ij} \bar f_{Li}  f_{Rj} H - i (\xi^f_A)_{ij} \bar f_{Li}  f_{Rj} A \right)\nonumber\\  &+ \frac{\sqrt{2}}{v} \sum_{k=1}^3 \bar u_{i} \left[ \left( m^u_i  (\xi^{u*}_A)_{ki}  V_{kj} P_L+ V_{ik}  (\xi^d_A)_{kj}  m^d_j P_R \right) \right] d_{j}  H^+ \nonumber\\  &+ \frac{\sqrt{2}}{v}  \bar \nu_i  (\xi^\ell_A)_{ij} m^\ell_j P_R \ell_j H^+ + H.c.\, \label{eq:Yukawa_CH}
\end{align} 
The reduced Yukawa couplings are given in Table~\ref{coupIII}, in terms of the free parameters $\chi_{ij}^f$ and the mixing angle $\alpha$ and  of $\tan\beta$. The terms proportional to $\delta_{ij}$ refers to the 2HDM Type-II whereas the term proportional to $\chi_{ij}^f$  denotes the new contribution of 2HDM Type-III.

Following the above arguments, the crucial elements of the Yukawa sector are then derived in terms of the $\chi_{ij}^f$'s. Clearly, these new free parameters are related to the masses and couplings of both quarks and leptons, consequently, one should be certain that rare decays, i.e., those that are suppressed in the SM, do not fail current bounds. Specifically, one should test the contributions of Higgs bosons to the relevant FCNC processes of $B$ mesons. To start with, the  off-diagonal terms are set to zero for simplicity. Furthermore, note that constraints from $\Delta B = 2$ processes are negligible due to the suppression factor $\sqrt{m^f_j m^f_i}/v$. However, transitions involving $\Delta B = 1$ processes are  taken into account in the analysis. Similarly, the $b\to s\gamma$  loop transition is sensitive to new physics, as any differences between the currently measured value of its rate and SM predictions may be interpreted in the form of a light charged Higgs boson with appropriate Yukawa couplings. 
Since no CP violation is observed in the lepton sector, it is reasonable to assume that $\chi_{ij}$ are real numbers the matrix is symmetric. These parameters may affect flavor changing neutral current in the Higgs sector $h\to l_i \bar{l_j}$  \cite{Diaz-Cruz:2004wsi, Benbrik:2015evd} which is proportional to non-diagonal elements. Once such effects, directly connected to the structure of the texture, have been taken care of, though, it is just matter of testing the Higgs sector of the 2HDM Type-III  against both theoretical and experimental conditions, which we do in the next section.
\begin{table*}
	\begin{center}
		\setlength{\tabcolsep}{13pt}
		\renewcommand{\arraystretch}{0.8} %
		\begin{tabular}{c|c|c|c} \hline\hline 
			$\phi$  & $(\xi^u_{\phi})_{ij}$ &  $(\xi^d_{\phi})_{ij}$ &  $(\xi^\ell_{\phi})_{ij}$  \\   \hline
			$h$~ 
			& ~ $  \frac{c_\alpha}{s_\beta} \delta_{ij} -  \frac{c_{\beta-\alpha}}{\sqrt{2}s_\beta}  \sqrt{\frac{m^u_i}{m^u_j}} \chi^u_{ij}$~
			& ~ $ -\frac{s_\alpha}{c_\beta} \delta_{ij} +  \frac{c_{\beta-\alpha}}{\sqrt{2}c_\beta} \sqrt{\frac{m^d_i}{m^d_j}}\chi^d_{ij}$~
			& ~ $ -\frac{s_\alpha}{c_\beta} \delta_{ij} + \frac{c_{\beta-\alpha}}{\sqrt{2}c_\beta} \sqrt{\frac{m^\ell_i}{m^\ell_j}}  \chi^\ell_{ij}$ ~ \\
			$H$~
			& $ \frac{s_\alpha}{s_\beta} \delta_{ij} + \frac{s_{\beta-\alpha}}{\sqrt{2}s_\beta} \sqrt{\frac{m^u_i}{m^u_j}} \chi^u_{ij} $
			& $ \frac{c_\alpha}{c_\beta} \delta_{ij} - \frac{s_{\beta-\alpha}}{\sqrt{2}c_\beta} \sqrt{\frac{m^d_i}{m^d_j}}\chi^d_{ij} $ 
			& $ \frac{c_\alpha}{c_\beta} \delta_{ij} -  \frac{s_{\beta-\alpha}}{\sqrt{2}c_\beta} \sqrt{\frac{m^\ell_i}{m^\ell_j}}  \chi^\ell_{ij}$ \\
			$A$~  
			& $ \frac{1}{t_\beta} \delta_{ij}- \frac{1}{\sqrt{2}s_\beta} \sqrt{\frac{m^u_i}{m^u_j}} \chi^u_{ij} $  
			& $ t_\beta \delta_{ij} - \frac{1}{\sqrt{2}c_\beta} \sqrt{\frac{m^d_i}{m^d_j}}\chi^d_{ij}$  
			& $t_\beta \delta_{ij} -  \frac{1}{\sqrt{2}c_\beta} \sqrt{\frac{m^\ell_i}{m^\ell_j}}  \chi^\ell_{ij}$ \\ \hline \hline 
		\end{tabular}
	\end{center}
	\caption {Yukawa couplings of the $h$, $H$, and $A$ bosons to the quarks and leptons in the 2HDM Type-III.} 
	\label{coupIII}
\end{table*}

\section{ Explanation of the 96 GeV Excesses at LEP and CMS}
In this section, we investigate whether the 2HDM Type-III can explain simultaneously the 96 GeV excesses observed by
both LEP (across ADLO) and LHC (CMS, in particular) experiments. We follow  the approach to explain these excesses  already adopted  in  \cite{Cao:2016uwt,Heinemeyer:2021msz,Biekotter:2019kde,Biekotter:2021qbc} which defines the LEP and CMS signal strengths in terms of production cross section ($\sigma$) and decay Branching Ratios ({\cal BR}s) as follows:
\begin{eqnarray}
	\mu_{\mathrm{LEP}}^{\mathrm{bb}}&=&\frac{\sigma_{\rm 2HDM}(e^+e^-\to Zh)}{\sigma_{\rm SM}(e^+e^-\to Zh)}\times \frac{{\cal BR}_{\rm 2HDM}(h\to b\bar{b})}{{\cal BR}_{\rm SM}(h_{\rm SM}\to b\bar{b})}\nonumber =\left|c_{hZZ}\right|^2\times \frac{{\cal BR}_{\rm 2HDM}(h\to b\bar{b})}{{\cal BR}_{\rm SM}(h_{\rm SM}\to b\bar{b})},\label{mu_lep}
\end{eqnarray} 	
\begin{eqnarray}
	\mu_{\rm CMS}^{\mathrm{\gamma\gamma}}&=&\frac{\sigma_{\rm 2HDM}(gg\to h)}{\sigma_{\rm SM}(gg\to h_{\rm SM})}\times \frac{{\cal BR}_{\rm 2HDM}(h\to \gamma\gamma)}{{\cal BR}_{\rm SM}(h_{\rm SM}\to \gamma\gamma)}\nonumber =\left|c_{htt}\right|^2\times \frac{{\cal BR}_{\rm 2HDM}(h\to \gamma\gamma)}{{\cal BR}_{\rm SM}(h_{\rm SM}\to \gamma\gamma)},\label{mu_cms}
\end{eqnarray} 	
{\color{black} The $e^+ e^- \to Z h$ process receives large corrections from the charged Higgs scalar \cite{Xie:2018yiv}. Based on the charged Higgs mass and $\tan\beta$, the complete electroweak contributions to the standard one-loop model ranged from $-1.6\%$ to $+5.6\%$ considering different values of$m^2_{12}$. Such corrections may be transferred to $\left|c_{hZZ}\right|^2$. Moreover, $h\to b \bar{b}$ has been estimated at NLO in the overall 2HDM \cite{Xie:2018yiv, Kanemura:2017gbi}. We have checked nearly for $m_h \ge 90$ GeV, tree-level results and NLO are in pretty accurate agreement}. In addition,   $c_{htt}$ is the effective couplings of the light Higgs  top quarks, computed by \texttt{2HDMC}\cite{2HDMC}. Here, $h_{\rm SM}$ is the SM Higgs-boson with mass rescaled to that of  the lightest CP-even Higgs-boson $h$ of the 2HDM Type-III. We then evaluate such signal strengths using the experimental measurements 
\begin{eqnarray}
\mu_{\rm CMS}^{\mathrm{\gamma\gamma, exp}} = 0.6 \pm 0.2 \quad \quad\quad  \mu_{\mathrm{LEP}}^{\mathrm{bb, exp}} = 0.117 \pm 0.057
\end{eqnarray}
and performed a constrained (i.e., $m_h \sim 96$ GeV) to extract the ”best-fit” point by computing the smallest $\chi^2_{96}$, the latter given by
\begin{eqnarray}
	\chi^2_{96}=\left(\frac{\mu_{\mathrm{LEP}}^{\mathrm{bb}}-0.117}{0.057}\right)^2+\left(\frac{\mu_{\rm CMS}^{\mathrm{\gamma\gamma}}-0.6}{0.2}\right)^2.
\end{eqnarray}

\section{Theoretical and Experimental Bounds}
\label{sec-A}
Before proceeding to show our results, we describe how we have constrained the parameter space of the 2HDM Type-III scenario that we are targeting, by using both theoretical requirements of self-consistency and experimental measurements from past and present machines. 

The generic 2HDM scalar potential is required to satisfy several theoretical conditions, in order to obtain a realistic and viable model. The theoretical requirements  are perturbativity of the scalar quartic couplings, vacuum stability and the tree-level perturbative unitarity conditions for various scattering amplitudes of gauge and Higgs boson states. These constraints map onto specific conditions of the parameter space of the model, as follows.
\begin{itemize}
	\item \textbf{Unitarity} constraints require  a variety of scattering process to be unitary: specifically, the tree-level 2-to-2 body scattering matrix involving scalar-scalar, gauge–gauge and/or scalar-gauge initial and/or final states must have eigenvalues $e_i$'s such that $|e_i|< 8\pi$ ~\cite{uni1,uni2,uni3}.
	\item \textbf{Perturbativity} constraints impose the following condition on the quartic couplings of the scalar potential:$|\lambda_i|<8\pi$ ~\cite{Branco:2011iw}.	
	\item \textbf{Vacuum stability} constraints require the potential  be bounded from below and positive in any direction of the fields $\Phi_i$, consequently, the parameter space must satisfy the following conditions~\cite{Barroso:2013awa,sta}:
	\begin{align}
		\lambda_1 > 0,\quad\lambda_2>0, \quad\lambda_3>-\sqrt{\lambda_1\lambda_2} ,\nonumber\\ \lambda_3+\lambda_4-|\lambda_5|>-\sqrt{\lambda_1\lambda_2}.\hspace{0.5cm}
	\end{align}
\end{itemize}

	Furthermore, on the experimental side, we consider high precision constraints arising from Electro-Weak Precision Observables (EWPOs),  measurements at the LHC of the properties of the newly discovered Higgs boson, null Higgs boson searches at  LEP, Tevatron and the LHC as well as flavour observables. 
\begin{itemize}
	\item \textbf{EWPOs}, implemented through the EW oblique parameters $S, T, U$ \cite{Grimus:2007if,oblique2}, require a  $95\%$ Confidence Level (C.L.) in matching the  global fit results~\cite{particle2020review}:
	\begin{align}
		S = -0.01 \pm 0.10,\quad T = 0.03 \pm 0.12,\nonumber\quad  U = 0.02 \pm 0.11.\hspace{1.5cm} 
	\end{align}
\item {\bf SM-like Higgs boson discovery}: an agreement between selected points in parameter space and the current measurements of the properties of the discovered Higgs boson at 125 GeV is enforced by means of the publicly available code \texttt{HiggsSignals-2.6.1} \cite{Bechtle:2008jh,Bechtle:2011sb,Bechtle:2013wla,Bechtle:2014ewa,Bechtle:2015pma,Bechtle:2020pkv,HS1,HS}, which computes a $\chi^2$ function by comparing the predictions of the model with the Higgs signal strengths from both Tevatron (marginally) and the LHC (crucially). In our analysis, we adopt $\Delta\chi^2_{125}$ = $\Delta\chi^2=\chi^2-\chi^2_{\mathring{\rm min}}$ to judge the validity of our generated points away from the ``best fit'' case at its minimum.

\item {\bf Non-SM-like Higgs boson exclusions}: 
to check the parameter space points against the exclusion limits from null Higgs boson searches at LEP, Tevatron and, in particular, the LHC, we apply the public code \texttt{HiggsBounds-5.10.1}  ~\cite{Bechtle:2008jh,Bechtle:2011sb,Bechtle:2013wla,Bechtle:2014ewa,Bechtle:2015pma,Bechtle:2020pkv}. 

\item
{\bf $B$-physics 
 observables}  are tested against data by resorting to the public code \texttt{SuperIso\_v4.1} \cite{superIso} and the experimental  measurements used are as follows:
	\begin{enumerate}
		\item ${\cal BR}(\overline{B}\to X_s\gamma)|_{E_\gamma<1.6\mathrm{~GeV}}=\left(3.32\pm0.3\right)\times 10^{-4}$~\cite{Amhis:2014hma,Bphy1},
		\item  ${\cal BR}(B_s\to \mu^+\mu^-)=\left(3.1\pm1.4\right)\times 10^{-9}$~\cite{Archilli:2014cla,Bphy2},
		\item ${\cal BR}(B^+\to \tau^+\nu_\tau)=1.06^{+0.38}_{-0.28}\times 10^{-4}$~\cite{Amhis:2014hma,Bphy3},
		\item $\frac{{\cal BR}(K \to \mu \nu)}{{\cal BR}(\pi \to \mu \nu)}=0.6358\pm0.0011$~\cite{ParticleDataGroup:2008zun,FlaviaNetWorkingGrouponKaonDecays:2008hpm},
	\end{enumerate} 

\item {Anomalous magnetic moment of the muon}, which is taken into account by requiring $\delta a_{\mu}=(2.95\pm0.88)\times10^{-9}$~\cite{Miller:2007kk}. In fact, the full $\delta a_\mu$ include all one-loop contributions and all two-loop Barr-Zee-type contributions. As we may see later, the parameters  $\chi^l_{11}$ and $\chi^l_{33}$  enter via the two-loop photon Barr-Zee diagram contribution from Higgs with all lepton running in the second loop \cite{Chang:2000ii,Cheung:2001hz}
\end{itemize}

\section{Numerical Results}

In line with the above sections, we consider $H$ to be the SM-like Higgs particle discovered at the LHC in 2012 and  $m_H$ = 125 GeV its mass,  therefore $m_h$ would be smaller (in what is normally referred to an ``inverted hierarchy'' scenario). We then perform a systematic scan over the 2HDM Type-III parameter space, as shown in eq.~(\ref{parm}). Given that we take $m_H$ = 125 GeV, the $S, T$ and $U$ constraints force the whole Higgs boson spectrum to be rather light. Specifically,    
the charged Higgs boson mass is taken in the range 140--170 GeV while the CP-odd is assumed to be in the range 70--90 GeV. With the assumption that $H$ is the well-known Higgs particle, taking into account all the LHC data will force the couplings of such state to the SM particles to be very similar to those predicted by the SM. Consequently, the coupling of the $H$ state to the $W^+W^-$ (and $ZZ$) gauge bosons, which is proportional to $\cos(\beta - \alpha)$, would be SM-like if $\cos(\beta -\alpha) \in [0.76, 0.83] $ within 1$\sigma$ level. Bearing all this in mind, we then perform a systematic numerical scan over the 2HDM Type-III parameter space as illustrated below:
\begin{eqnarray}
	\centering
	&&\hspace{0.7cm} m_{h} \in [80,~110]\ \text{GeV}, \ \ m_{H}= 125\ \text{GeV},\nonumber \\
	&& \sin(\beta-\alpha)\in [-0.5,~-0.1], \ \  m_{A}\in [70,~90]\ \text{GeV}, \nonumber \\
	&&\hspace{0.7cm} m_{H^\pm}\in [140,~180]\ \text{GeV}, \ \  \tan\beta\in [1.1,~1.5],\ \nonumber \\
	&& \hspace{1.5cm}m_{12}^2 = m_h^2\tan\beta/(1+\tan^2\beta).
	 \label{parm}
\end{eqnarray}
Of all constraints enforced, a particular role is played by the flavour ones from the $B$-meson sector, as these strongly 
limit the region of validity in the $(m_{H^\pm},\tan\beta)$  (see Fig. \ref{fig1}). From here, one can read that  a light charged Higgs boson (with mass from around $m_H$ and below $m_t$) is allowed, so long that $\tan \beta > 1.08$ (unlike the 2HDM Type-II, wherein the lower bound on charged Higgs mass at 95\% C.L. is 580 GeV or upwards, quite irrespectively of $\tan\beta$~\cite{Misiak:2017bgg}). This is instrumental to enable the 2HDM Type-III pursued here to also comply with the
EWPO constraints, as the charged Higgs boson is sufficiently degenerate with the neutral Higgs state, a precondition for
$S,T$ and $U$ to not fall foul of their experimental measurements.

Given the role played by the  $\chi_{ij}^f$   free parameters in the definition of the texture, hence on the Yukawa structure of the
2HDM Type-III studied here, we show in Fig. \ref{fig2} the correlation among such quantities in presence of all aforementioned flavour constraints (from both the quark and lepton sector), i.e.,  $B_{s,d}^0 \to \mu^+ \mu^-$, $\overline{B} \to X_s \gamma$,  $\frac{{\cal BR}(K \to \mu \nu)}{{\cal BR}(\pi \to \mu \nu)}$ and $\delta a_{\mu}$. From investigating such a figure, we note that 
the on-diagonal matrix elements in the up and down sectors are not as large as in the leptonic sector. It is also interesting to note  that $\chi_{33}^{u,d}$ are both strongly sensitive to the $B_{s,d}^0 \to \mu^+ \mu^-$ and $\overline{B} \to X_s \gamma$ processes.  The best fit to all such flavour observables is given by the diamond symbol, which corresponds to the following configuration:
\begin{figure}[t!]	
	\centering
	\includegraphics[height=6.2cm,width=8cm]{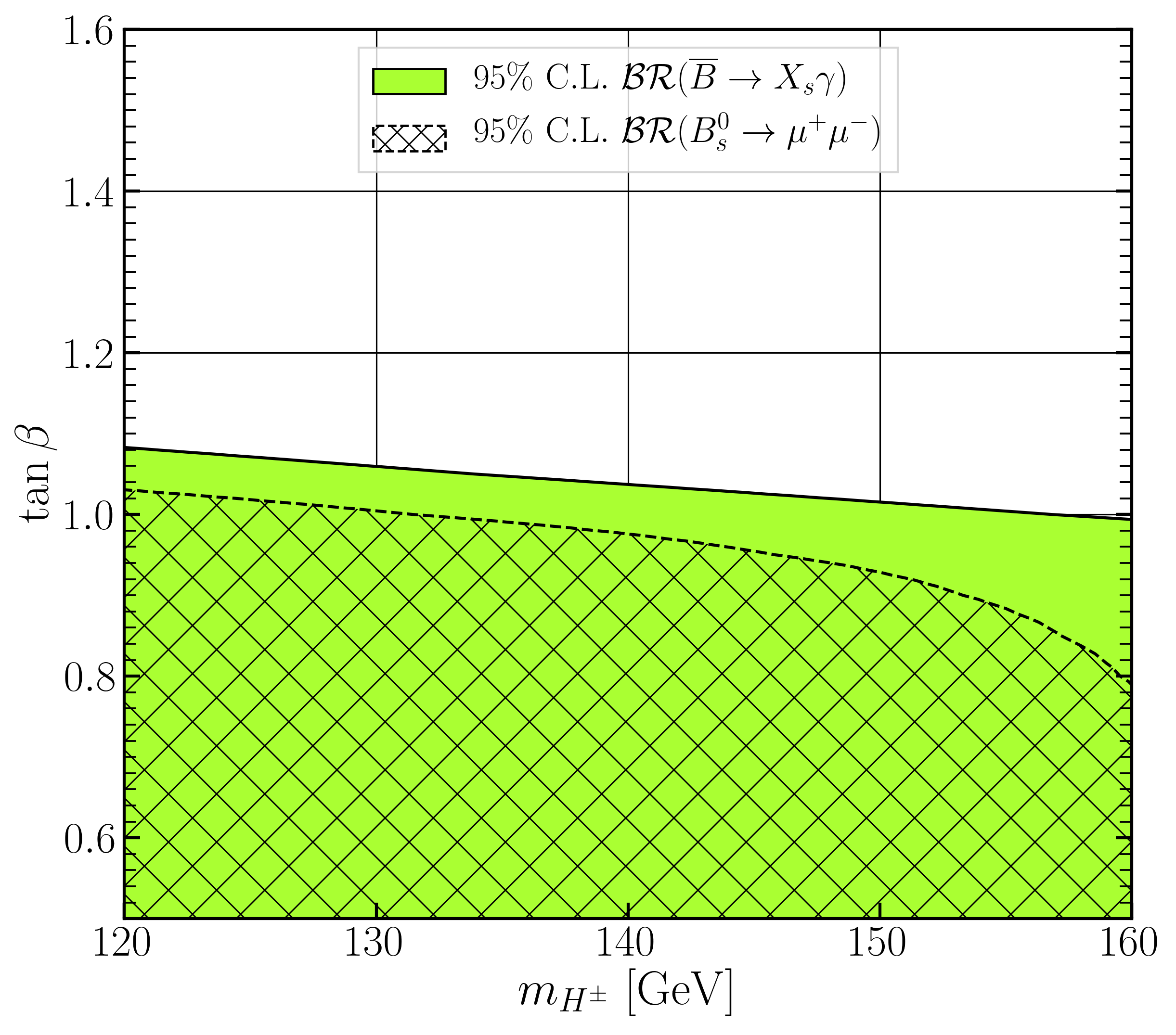}	
	\caption{Excluded regions on the $(m_{H^\pm},~\tan\beta)$ parameter space by flavour constraints at $95\%$ C.L.}\label{fig1}
\end{figure}
\begin{figure}[b!]	
	\centering
	\includegraphics[height=17cm,width=16cm]{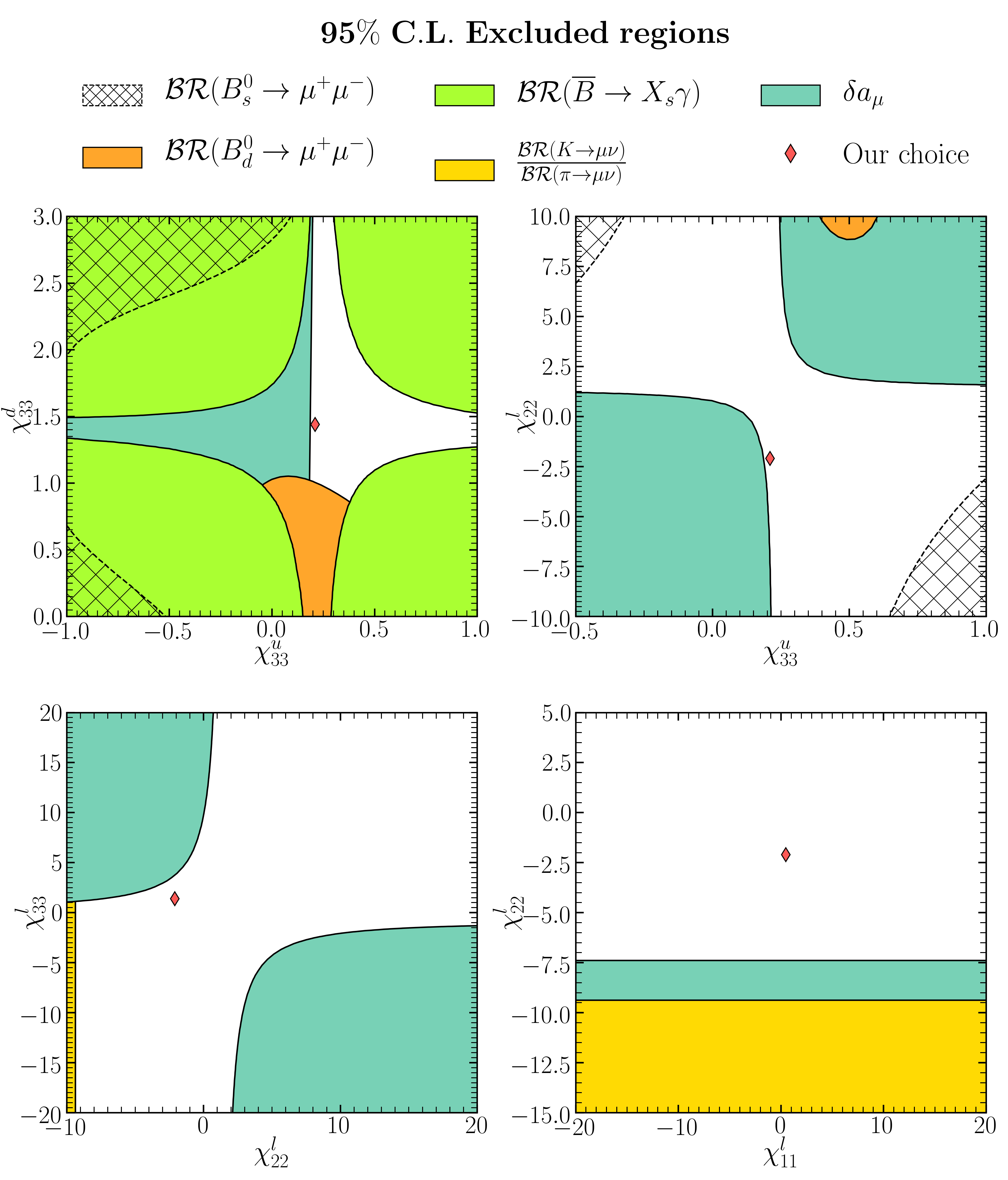}
	\caption{Excluded regions on the $\chi_{ij}$ parameter space  by flavour constraints at $95\%$ C.L.}\label{fig2}
\end{figure}

\begin{equation}
	\chi^u=\begin{pmatrix}
		0.187 & 0 & 0 \\
		0 & 0.254 & 0\\
		0 & 0 & 0.210
	\end{pmatrix},~
	\chi^d=\begin{pmatrix}
		-0.553 & 0 & 0 \\
		0 & 2.863 & 0\\
		0 & 0 & 1.440
	\end{pmatrix},~
	\chi^l=\begin{pmatrix}
		0.484 & 0 & 0 \\
		0 & -2.101 & 0\\
		0 & 0 & 1.400
	\end{pmatrix}.
	\label{chi_bestIII}
\end{equation}

To begin with, the LEP constraints on the Higgs strahlung process $e^+e^- \to Z^{(*)}h$ may restrict an important range of $\sin^2(\beta-\alpha)$ values for any given choice of $m_h$,  as long as the process is kinematically allowed. Because the $ZZh$ coupling in the 2HDM Type-III we are considering is suppressed by $\sin^2(\beta-\alpha)$ with respect to the SM value and considering (from LHC data) that, in our scenario, $\sin^2(\beta-\alpha)$ is expected to be small, the cross sections for $e^+e^- \to Z^{(*)}h$ that we obtain are roughly in agreement with the LEP experimental limits. We also mention that the Tevatron has  investigated such a light boson via the $p\bar p \to Vh$ ($V=Z,W^\pm$) process \cite{CDF:2013jor}. These limits, however, are considerably less severe than LEP ones. Obviously, the masses and couplings entering the $e^+e^-$ process are also indirectly influencing the SM-like Higgs boson data collected at the LHC, not only through the mixing between $H$ and $h$, but also via the $H^\pm$ effects in $h\to \gamma\gamma$ decay mode.

\begin{figure}[h!]	
	\centering
	\includegraphics[height=8cm,width=10cm]{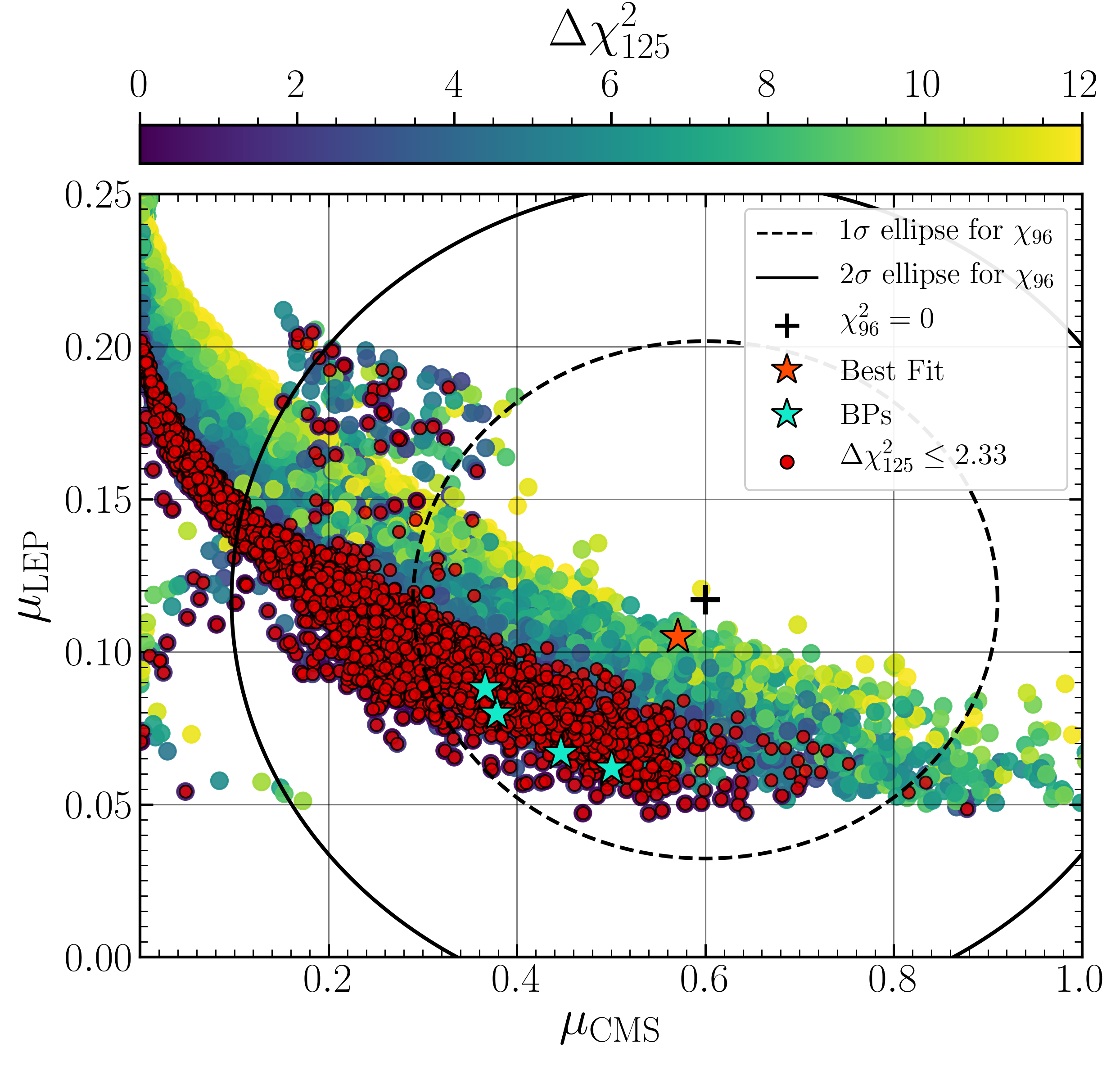}
	\caption{The signal strengths $\mu_{\rm CMS}$ and $\mu_{\mathrm{LEP}}$ following the scan described in the text. The dashed and solid black lines indicate the $1\sigma$ and $2\sigma$ ranges of $\chi_{96}$, respectively. The orange star corresponds to the best fit point 
in the $h$ mass range [94, 98] GeV. The colour code indicates the $\Delta\chi^2_{125}$ values.}\label{fig3}
\end{figure}

Fig.~\ref{fig3} shows the result of the described 2HDM Type-III scan over the ($\mu_{\rm CMS}, \mu_{\mathrm{LEP}}$) plane, where the colour code indicates $\Delta\chi^2_{125}$. The dashed and solid lines correspond to the 1$\sigma$ and 2$\sigma$ ellipses, respectively, and the orange star indicates the best fit point. All the points shown in the figure have $\Delta\chi^2_{125} \le 12$. An interesting observation is that there are many points (in red) with $\Delta \chi_{125}^2\le 2.33$  within the 1$\sigma$  ellipse. Also, our best fit point is near the center of the ellipses. Overall, the 2HDM Type-III  is fully capable of capturing the excess of 96 GeV. Finally, in this figure, we also identify as aqua stars four viable Benchmark Points (BPs) that we will use in the remainder of the paper.

\begin{figure}[h!]
	\centering
    \includegraphics[height=7cm,width=16.75cm]{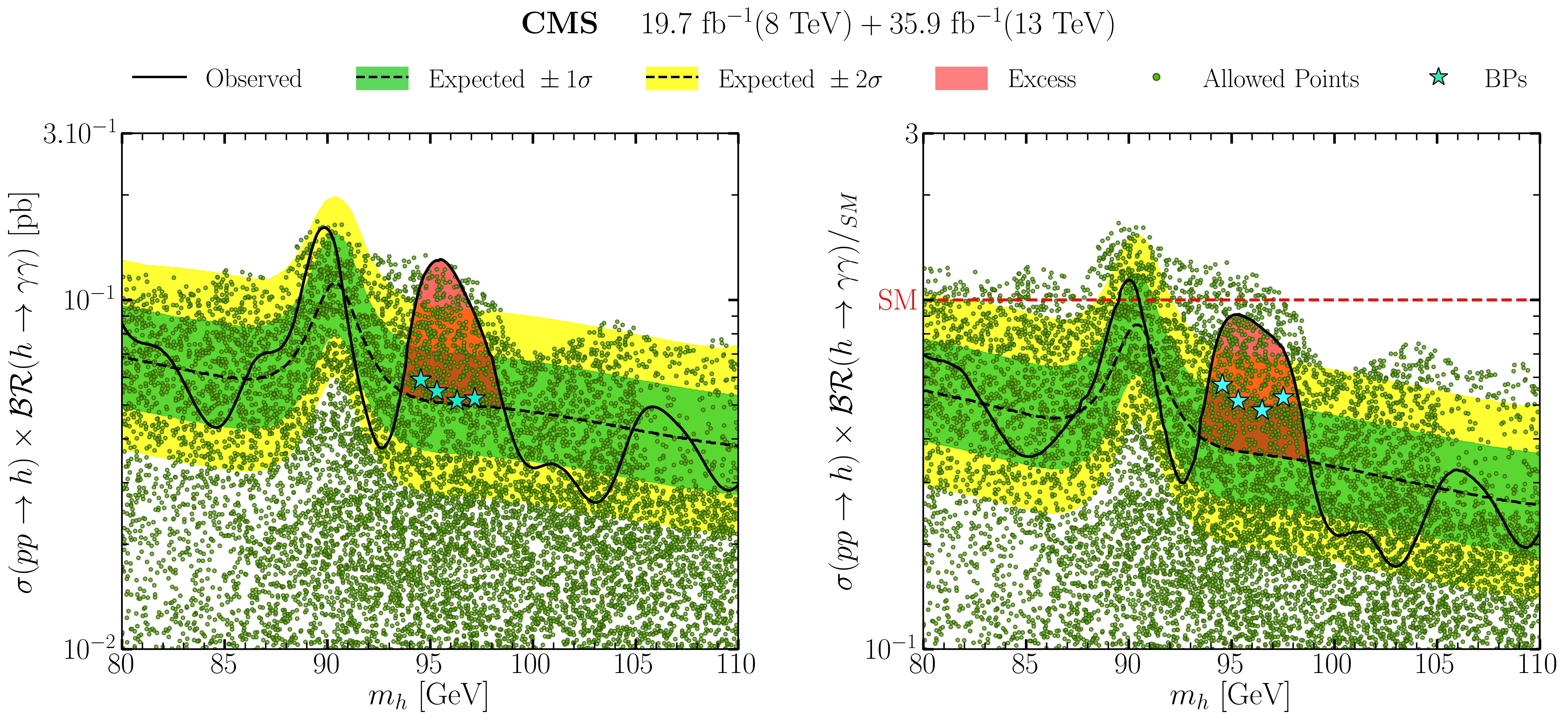}		
	\caption{Allowed points, following the discussed theoretical and experimental constraints,  
		superimposed on the results of the CMS $8+13$ TeV low-mass di-photon analysis \cite{CMS:2018cyk}. The dashed line corresponds to the expected upper limit on $\sigma \times {\cal BR}(h\rightarrow \gamma\gamma)$ (left) and $\sigma \times {\cal BR}(h\rightarrow \gamma\gamma)/\sigma^{SM} \times {\cal BR}^{SM}(h\rightarrow \gamma\gamma)$ (right) at $95\%$ C.L.,
		with $1$ and $2$ sigma errors in green and yellow, respectively. The solid line is the observed
		upper limit at $95\%$ C.L. The aqua stars represent the BPs already introduced.}\label{fig4}
\end{figure} 
\begin{figure}[h!]
	\centering
	\includegraphics[height=7cm,width=10cm]{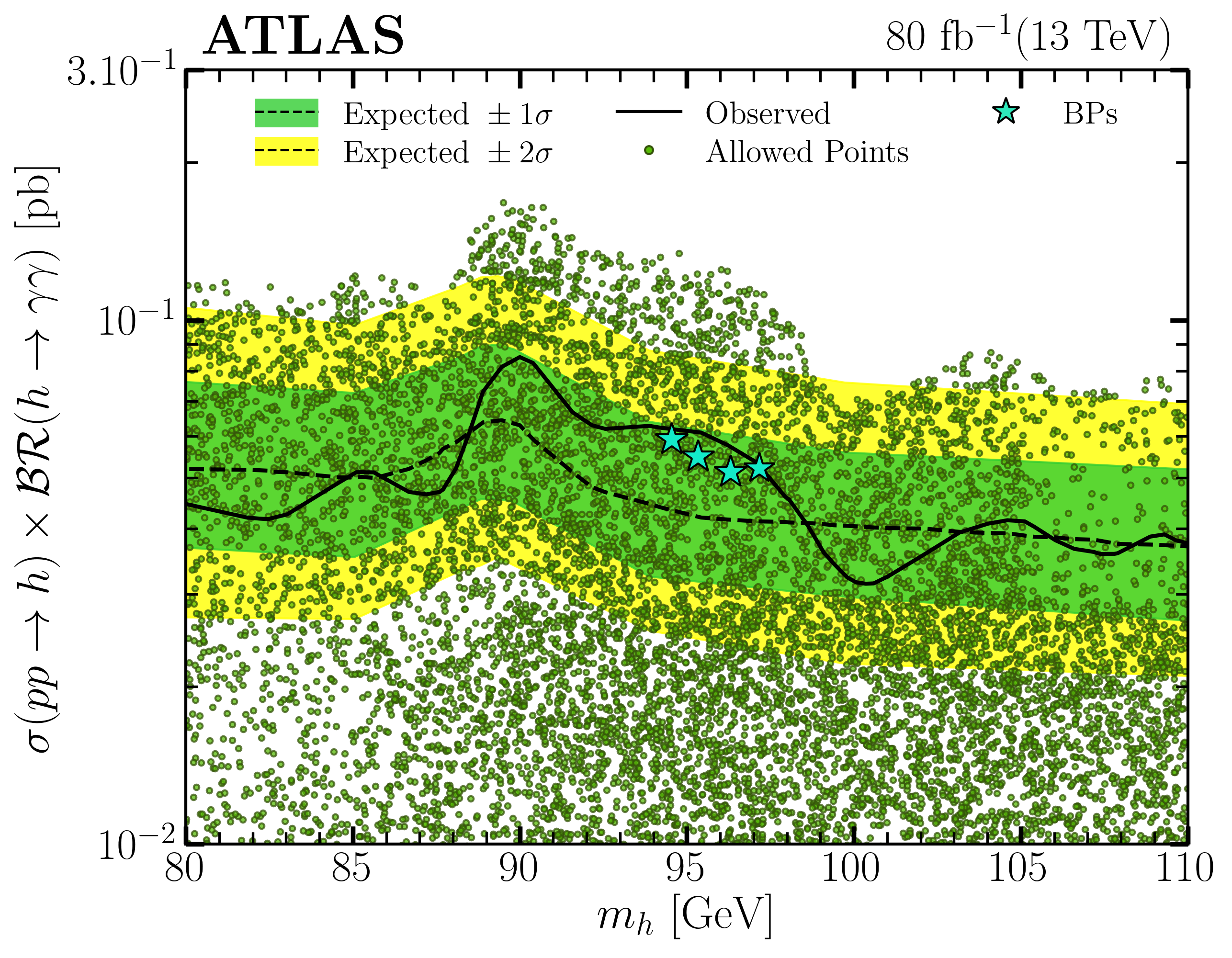}		
	\caption{Allowed points, following the discussed theoretical and experimental constraints,
		superimposed on the results of the ATLAS 13 TeV low-mass di-photon analysis \cite{ATLAS:2018xad}.
The meaning of the colours, lines and stars is the same as in the previous figure.}\label{fig5}
\end{figure} 

We show in Fig.~\ref{fig4} the predicted rate for $\sigma(pp \to h \to \gamma\gamma)$ (left) in our 2HDM Type-III and its ratio to
the SM results, 
$\sigma(pp \to h \to \gamma\gamma)/{\rm SM}$, for each parameter point, in combination with the expected and observed upper limits from the CMS
analysis \cite{CMS:2018cyk}. One can see that many points could indeed contribute to the excess observed by CMS in the $h \to \gamma\gamma$ final state. 
It is further interesting to note that this configuration of the parameter space is also tested against the exclusion limits given by the ATLAS Collaboration \cite{ATLAS:2018xad} on $\sigma \times {\cal BR}(h\to\gamma\gamma)$, captured by Fig.~\ref{fig5}. As one can see therein, the chosen BPs (amongst others) can account for the experimental results of both collaborations. 

\begin{figure}[H]
	\begin{minipage}{0.46\textwidth}
		\centering
		\includegraphics[height=7cm,width=8.3cm]{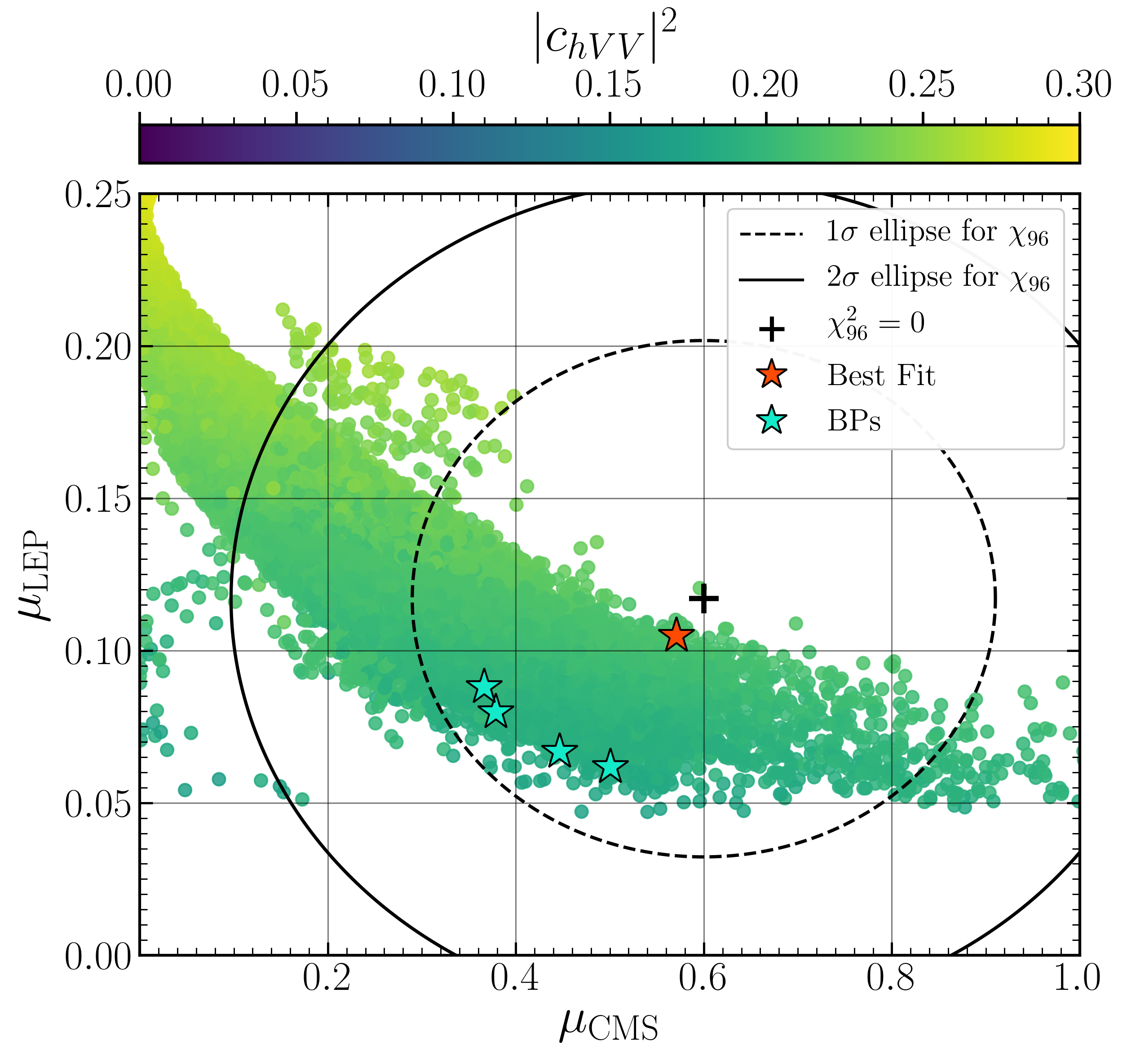}
	\end{minipage}\hspace{0.5cm}
	\begin{minipage}{0.46\textwidth}
		\centering
		\includegraphics[height=7cm,width=8.3cm]{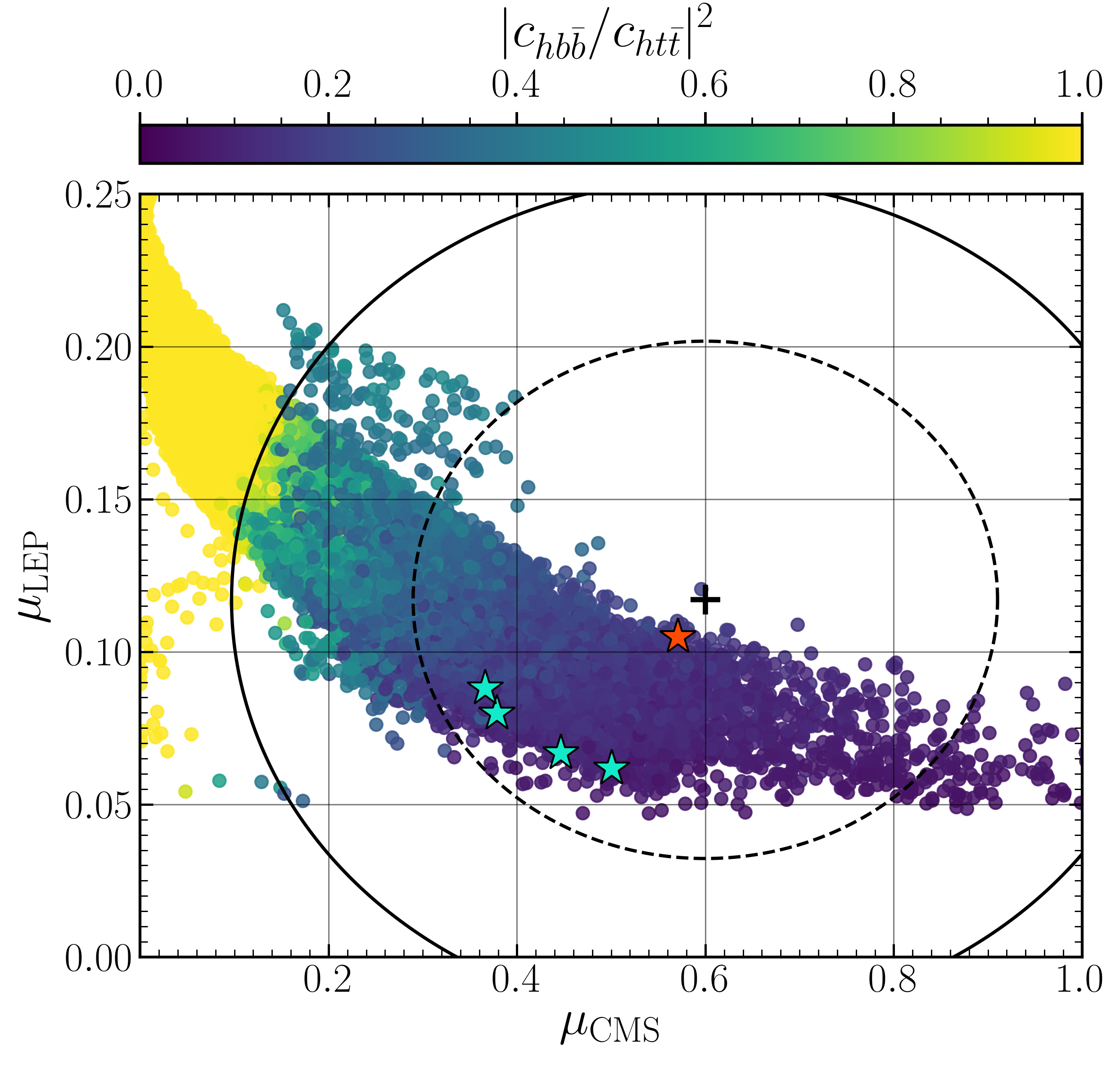}
	\end{minipage}			
	\caption{Same as in Fig.~\ref{fig3}, with the
		colour code indicating  $|c_{hVV}|^2$ (left)  and  $|c_{hb\bar{b}}/c_{ht\bar{t}}|^2$ 
(right).}\label{fig6}
\end{figure}

 
In Fig. \ref{fig6}, we show $|c_{hVV}|^2$ and $|c_{hb\bar{b}}/c_{ht\bar{t}}|^2$ on the ($\mu_{\rm CMS}$, $\mu_{\mathrm{LEP}}$) plane. It can be observed from the left panel that the points with higher signal strength $\mu_{\mathrm{LEP}}$ always have the higher coupling $c_{hVV}$, since  $\mu_{\mathrm{LEP}}$ is directly proportional to $|c_{hVV}|^2$ (see eq.~(\ref{mu_lep})). Furthermore, the points with lower values of $|c_{hb\bar{b}}/c_{ht\bar{t}}|^2$, in the right panel, give a higher $\mu_{\rm CMS}$ signal strength, as $\mu_{\rm CMS}$ is anti-proportional to $|c_{hb\bar{b}}/c_{ht\bar{t}}|^2$ (see eq.~(\ref{mu_cms})). However, a too low $c_{hb\bar{b}}$ coupling would  suppress ${\cal BR}(h \to b\bar{b})$ resulting in a smaller $\mu_{\mathrm{LEP}}$.

\begin{figure}[H]	
	\begin{minipage}{0.46\textwidth}
		\centering
		\includegraphics[height=7cm,width=8.3cm]{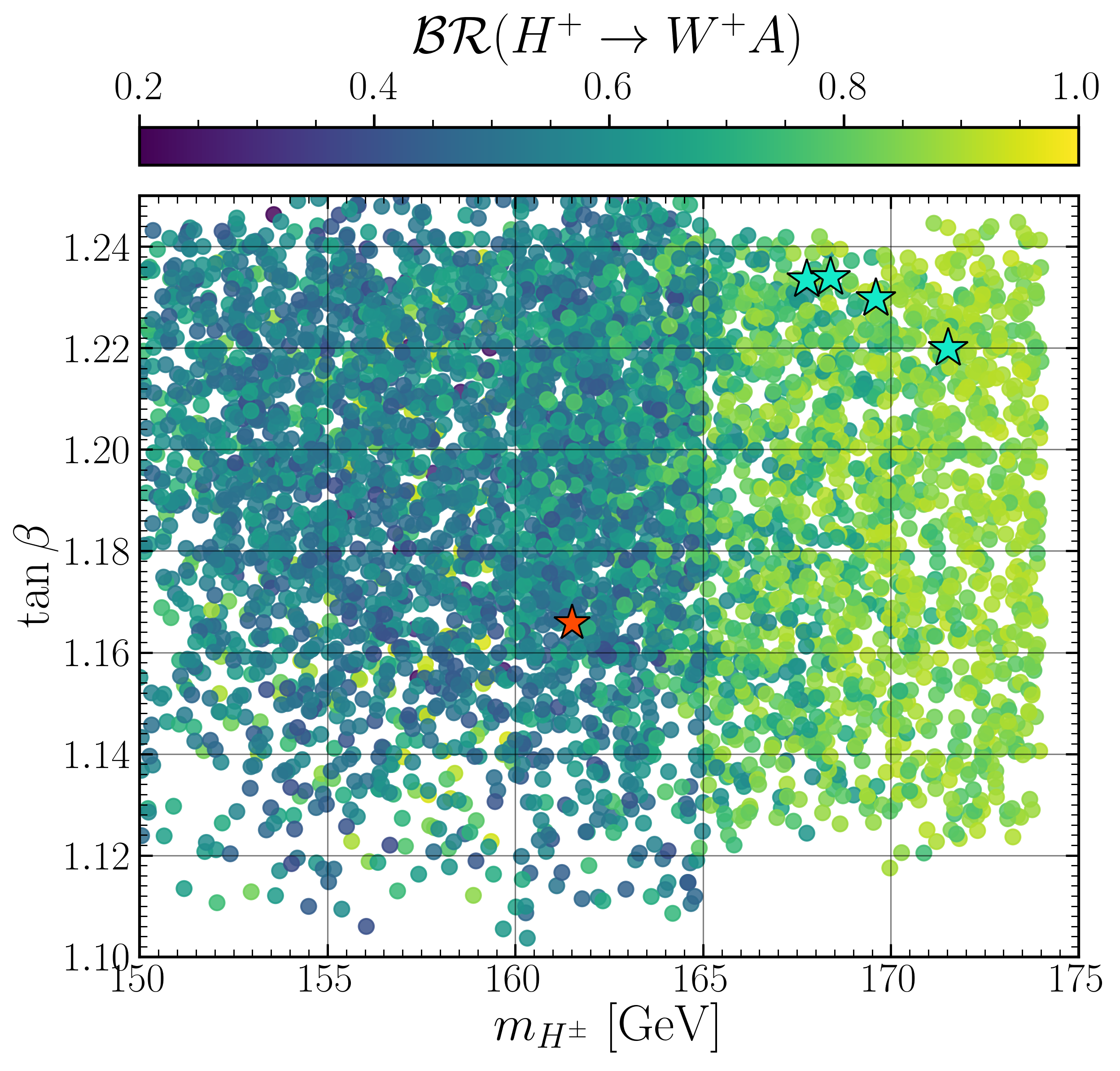}
	\end{minipage}\hspace{0.5cm}
	\begin{minipage}{0.46\textwidth}
		\centering
		\includegraphics[height=7cm,width=8.3cm]{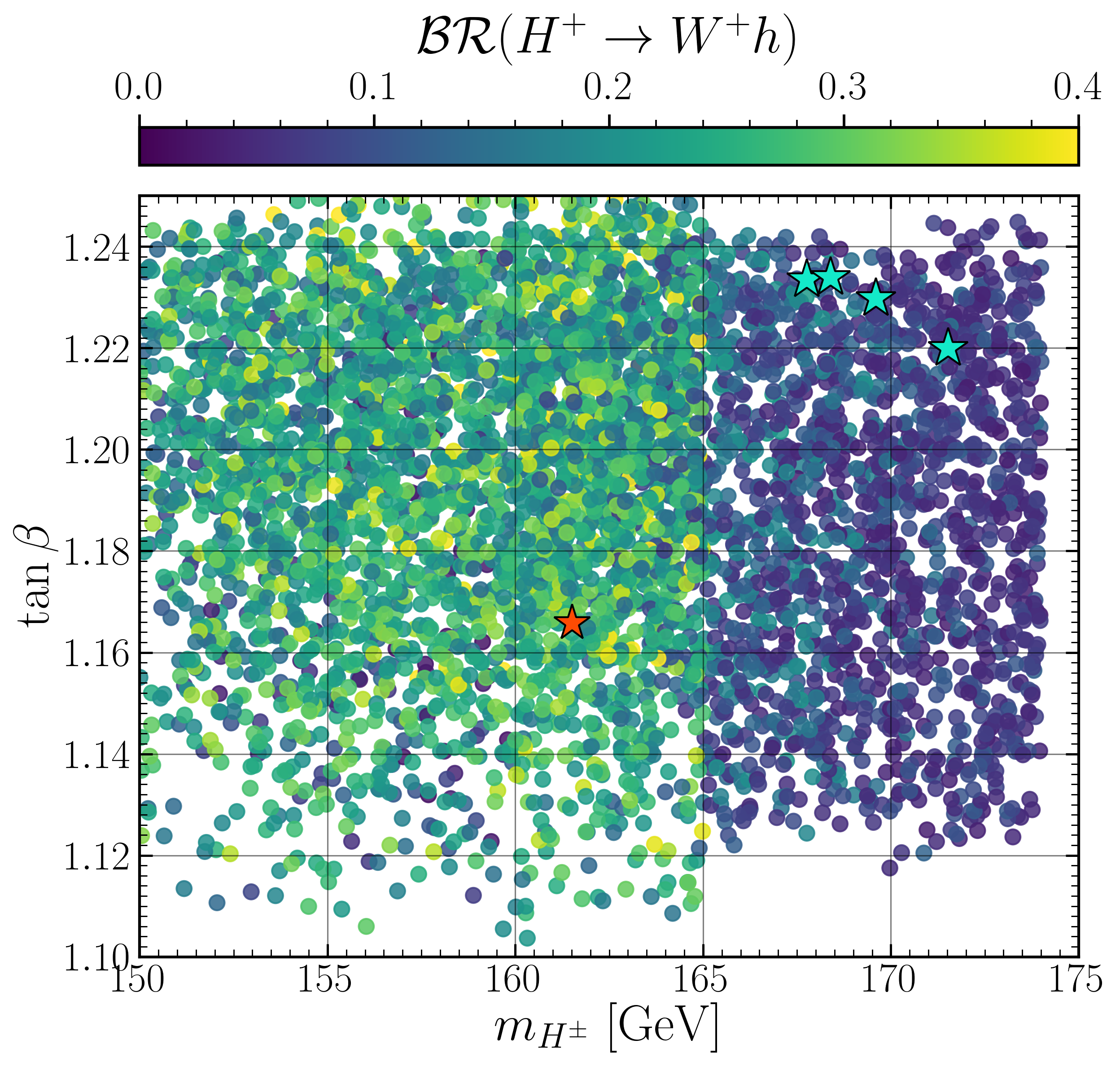}
	\end{minipage}
	\caption{The  allowed points within the 1$\sigma$ ellipse  on the $(m_{H^\pm}, \tan\beta)$ plane with the colour code indicating ${\cal BR}(H^+\to W^+A)$ (left) and ${\cal BR}(H^+\to W^+h)$ (right). Both the best fit (to theoretical and experimental constraints) point and the usual four BPs are also reported.}
	\label{fig7}	
\end{figure}

Having satisfied ourselves that the 2HDM Type-III scenario introduced here is able to explain the LEP and CMS anomalies, over
a specific parameter space region, we now turn to identifying a smoking gun signatures of it. We concentrate here on the $H^\pm$ state, that we have seen (recall Fig.~\ref{fig1}) can be rather light, given the fact that also $h$ and $A$ are light, so that this calls for testing the possibility of sizeable $H^+\to W^+ A$ and/or $H^+\to W^+ h$ decay rates.   
 For this purpose, 
in Fig.~\ref{fig7}, we present the allowed points, within the usual 1$\sigma$ ellipse, mapped against  ${\cal BR}(H^\pm\to W^+A)$ (left panel) and ${\cal BR}(H^+\to Wh)$ (right panel). One can read from this figure that the charged Higgs decay width is dominated
by the decay channel $W^+A$, for which the ${\cal BR}$ could reach 82\% for $m_{H^\pm}>165$~GeV. The decay 
channel $W^+h$, for which the ${\cal BR}$ could reach 38\% for $m_{H^\pm}<165$~GeV exceeding the fermonic decay modes of light charged Higgs $c\bar s$ and $\tau\nu$ and without conflict with the experimental limits. {The panels in Fig.~\ref{fig7} illustrate that  $ {\cal BR}(H^\pm \to  hW^\pm )  + {\cal BR}(H^\pm \to W^\pm A)   \sim 100\%$
 in the allowed parameter regions. As consequence, searches for these channels may act as a collateral signatures validating our 2HDM Type-III explanation of
the discussed excesses. }

Before closing, for the purpose of encouraging experimental analyses of this scenario at the LHC, chiefly, to both confirm our findings in the $\gamma\gamma$ channel and attempting to extract  the hallmark signatures of it, $H^+\to W^+A$ and/or $H^+\to W^+h$, in Tab.~\ref{Bp}, we present the four BPs in full, noting that they cover a range of light Higgs masses between 94 and 98 GeV, i.e., $\pm2$ GeV from the anomaly mass value seen by CMS, thereby consistent with the typical resolution of a CMS detector and also capturing 
the mass value of the LEP excess. {Recall that, in all BPs, the $H$ state is SM-like with $\Gamma(H) = 4.09$ MeV and its decay modes are within the current measurements. The cross section rates are also shown.}
\section{Conclusions}
In this paper, we have  illustrated how  excesses identified by the CMS and LEP Collaborations may potentially be ascribed to a Higgs boson produced by $gg$ fusion and decaying into $\gamma\gamma$ and $b\bar b$ with mass around 96 GeV, by adopting as theoretical framework a 2HDM Type-III with a suitable Yukawa texture. Under the assumption that the heaviest CP-even Higgs state, $H$, is the discovered one at the LHC in 2012, we have identified the regions of parameter space where the light CP-even state, $h$, can explain 
such excesses while being fully compliant with the  required signal strengths measured at both colliders.  This has been accomplished after applying standard theoretical requirements
of self-consistency as well as up-to-date  experimental constraints.

We  suggest that our results are compelling enough to call for a more thorough experimental examination of the 2HDM Type-III proposed here.
In this respect, we have highlighted the fact that the aforementioned excesses  occur in regions of  parameter space which can be probed by the LHC with current and future data
in smoking gun signatures involving a light charged Higgs state decaying into a $W^\pm$ boson and either the $h$ itself or the CP-odd companion, $A$. To facilitate  this, we have presented four BPs amenable to phenomenological investigation.

\color{black}
\noindent{\bf Acknowledgments}

\noindent
SM acknowledges support from the STFC Consolidated Grant ST/L000296/1 and is partially financed through the NExT Institute. The work of SS is supported in full by the NExT Institute. The work of RB and MB is supported by the Moroccan Ministry of Higher Education and Scientific Research MESRSFC and CNRST Project PPR/2015/6. 
\begin{table}[H]
	\begin{center}
		\setlength{\tabcolsep}{30pt}
		\renewcommand{\arraystretch}{0.8}
		\begin{adjustbox}{max width=\textwidth}		
			\begin{tabular}{lcccc}
				\hline\hline
				Parameters &       BP$1$ &       BP$2$ &       BP$3$ &       BP$4$  \\\hline\hline
				\multicolumn{5}{c}{(Masses are in GeV)} \\\hline
				$m_h$   &    94.55&    95.33 &    96.49 &    97.52 \\
				$m_H$   &   125 &   125 &   125 &   125 \\
				$m_A$   &    87.95 &    83.36&    85.13 &    82.27 \\
				$m_{H^\pm}$   &   167.76 &   168.39 &   169.60 &   171.52\\
				$\tan\beta$ &     1.23 &     1.23 &     1.22 &     1.21 \\
				$\sin(\beta-\alpha)$  &    $-0.420$ &    $-0.423$ &    $-0.431$ &   $ -0.438$ \\
				$\lambda_6$   &     0 &     0 &     0 &     0 \\
				$\lambda_7$   &     0 &     0 &     0 &     0 \\
				$m_{12}^2$     &  4373.63 &  4445.61 &  4557.44 &  4662.92 \\
				\hline\hline\multicolumn{5}{c}{Effective coupling $|c_{hXY}|$} \\\hline
				$|c_{hVV}|$ &  0.427854 &  0.430945 &  0.439079 &  0.446200 \\
				$|c_{hb\bar{b}}|$ &  0.046293 &  0.048631 &  0.054028 &  0.058105 \\
				$|c_{ht\bar{t}}|$ &  0.144724 &  0.140352 &  0.131938 &  0.127975 \\
				\hline\hline\multicolumn{5}{c}{Effective coupling $|c_{HXY}|$} \\\hline		
				$|c_{HVV}|$   &  0.922878 &  0.921439 &  0.917591 &  0.914149 \\
				$|c_{Hb\bar{b}}|$   &  0.664166 &  0.663986 &  0.663746 &  0.663866 \\
				$|c_{Ht\bar{t}}|$   &  1.167053 &  1.167399 &  1.169637 &  1.173210 \\
				$|c_{H\tau\tau}|$ &  1.068317 &  1.067894 &  1.067330 &  1.067330 \\
				\hline\hline\multicolumn{5}{c}{Collider signal strength} \\\hline
				$\mu_{\rm CMS}$ &  0.5 &  0.446 &  0.378&  0.366 \\
				$\mu_{\mathrm{LEP}}$ &  0.061 &  0.066 &  0.079 &  0.087 \\
				\hline\hline\multicolumn{5}{c}{Total decay width in GeV} \\\hline
				$\Gamma(h)$   &  0.000030 &  0.000031 &  0.000034 &  0.000037 \\
				$\Gamma(H)$  &  0.004099 &  0.004099 &  0.004092 &  0.004094 \\
				$\Gamma(A)$  &  0.000432 &  0.000401 &  0.000415 &  0.000402 \\
				$\Gamma(H^{\pm})$ &  0.010770 &  0.030775 &  0.028097 &  0.063084 \\
				\hline\hline	\multicolumn{5}{c}{${\cal BR}(h\to XY)$ in \%} \\\hline
				
				${\cal BR}(h\to \gamma\gamma)$   &   3.4553 &   3.4289 &   3.3538 &   3.2808 \\
				${\cal BR}(h\to gg)$     &  11.5384 &  10.6915 &   9.0740 &   8.1683 \\
				${\cal BR}(h\to b\bar{b})$     &  27.2348 &  28.9708 &  33.2201 &  35.6635 \\
				${\cal BR}(h\to c\bar{c})$     &   6.5675 &   5.8372 &   4.5686 &   3.8995 \\
				${\cal BR}(h\to s\bar{s})$     &   4.9132 &   4.6901 &   4.2159 &   3.7984 \\
				${\cal BR}(h\to\mu^+\mu^-)$   &  31.2203 &  30.0817 &  27.6151 &  25.3039 \\
				${\cal BR}(h\to\tau^+\tau^-)$ &   8.3852 &   8.6082 &   9.1459 &   9.3359 \\
				${\cal BR}(h\to ZZ)$      &   0.8641 &   0.9105 &   0.9797 &   1.0440 \\
				${\cal BR}(h\to W^+W^-)$     &   5.7995 &   6.4856 &   7.6074 &   8.7173 \\
				
				\hline\hline	\multicolumn{5}{c}{${\cal BR}(H\to XY)$ in \%} \\\hline
				${\cal BR}(H\to \gamma\gamma)$   &   0.1457 &   0.1452 &   0.1436 &   0.1418 \\
				${\cal BR}(H\to gg)$     &  10.5405 &  10.5478 &  10.6082 &  10.6697 \\
				${\cal BR}(H\to b\bar{b})$     &  55.6194 &  55.5916 &  55.6499 &  55.6419 \\
				${\cal BR}(H\to c\bar{c})$     &  7.1916 &  7.1947 &  7.2305 &  7.2670 \\
				${\cal BR}(H\to \tau^+\tau^-)$ &   6.9585 &   6.9530 &   6.9559 &   6.9518 \\
				${\cal BR}(H\to ZZ)$     &   2.1415 &   2.1349 &   2.1206 &   2.1035 \\
				${\cal BR}(H\to W^+W^-)$     &  17.0929 &  17.0404 &  16.9266 &  16.7901 \\
				
				\hline\hline	\multicolumn{5}{c}{${\cal BR}(A\to XY)$ in \%} \\\hline
				${\cal BR}(A\to \gamma\gamma)$   &   0.0498 &   0.0454 &   0.0472 &   0.0447 \\
				${\cal BR}(A\to gg)$     &  29.6754 &  27.6545 &  28.4896 &  27.3641 \\
				${\cal BR}(A\to c\bar{c})$     &  13.1940 &  13.5884 &  13.4690 &  13.8044 \\
				${\cal BR}(A\to b\bar{b})$     &  49.6664 &  51.1480 &  50.5130 &  51.2798 \\
				${\cal BR}(A\to s\bar{s})$     &   0.6568 &   0.6773 &   0.6641 &   0.6633 \\
				${\cal BR}(A\to \mu\mu)$   &   1.9262 &   1.9648 &   1.9301 &   1.9028 \\
				${\cal BR}(A\to \tau\tau)$ &   4.8315 &   4.9217 &   4.8870 &   4.9409 \\
				
				\hline\hline\multicolumn{5}{c}{${\cal BR}(H^{\pm}\to XY)$ in \%} \\\hline
				${\cal BR}(H^{\pm}\to\mu\nu)$ &   0.1473 &   0.0518 &   0.0568 &   0.0253 \\
				${\cal BR}(H^{\pm}\to cs)$   &   0.8415 &   0.2951 &   0.3277 &   0.1501 \\
				${\cal BR}(H^{\pm}\to W^+h)$   &  21.3231 &   7.3599 &   8.0904 &   3.9528 \\
				${\cal BR}(H^{\pm}\to W^+A)$   &  59.6370 &  85.5962 &  83.2112 &  91.2625 \\
				${\cal BR}(H^{\pm}\to\tau\nu)$ &   0.3697 &   0.1298 &   0.1440 &   0.0657 \\
				${\cal BR}(H^{\pm}\to tb)$   &  17.3572 &   6.4452 &   8.0145 &   4.4573 \\
				\hline\hline\multicolumn{5}{c}{$\sigma$ in [pb]} \\\hline
				$\sigma(pp\to h\to \gamma\gamma)$ & 0.0591 &  0.0548 &  0.0513 &  0.0522 \\
				$\sigma(pp\to h\to \gamma\gamma)/{\rm SM}$ & 0.5715 &  0.5135 &  0.4825 &  0.5257  \\\hline\hline
				
			\end{tabular}
		\end{adjustbox}
	\end{center}
	\caption{The full description of our four BPs.}\label{Bp}
\end{table}




\begin{thebibliography}{50}
	\footnotesize
	\bibitem{Aad:2012tfa}
	ATLAS Collaboration, G.~Aad \underline{et al.}, ``Observation of a new particle in the search for the Standard Model Higgs boson with the ATLAS detector at the LHC'', \href{https://www.sciencedirect.com/science/article/pii/S037026931200857X}{\underline{Phys. Lett.} \textbf{B716} (2012) 1-29}, \href{https://arxiv.org/pdf/1207.7214.pdf}{\texttt{arXiv:1207.7214 [hep-ex]}}.\vspace{0.1cm}	
	
	\bibitem{Chatrchyan:2012ufa}
	CMS Collaboration, S.~Chatrchyan \underline{et al.}, ``Observation of a New Boson at a Mass of 125 GeV with the CMS Experiment at the LHC'', \href{https://www.sciencedirect.com/science/article/pii/S0370269312008581}{\underline{Phys. Lett.} \textbf{B716} (2012) 30-61}, \href{https://arxiv.org/pdf/1207.7235.pdf}{\texttt{arXiv:1207.7235 [hep-ex]}}.\vspace{0.1cm}
	
	\bibitem{Gunion:1992hs}
	J.~F.~Gunion, H.~E.~Haber, G.~L.~Kane and S.~Dawson, ``Errata for the Higgs hunter's guide'', \href{https://arxiv.org/pdf/hep-ph/9302272.pdf}{\texttt{arXiv:hep-ph/9302272 [hep-ph]}}.\vspace{0.1cm}
	
	\bibitem{Branco:2011iw} 
	G.~C.~Branco, P.~M.~Ferreira, L.~Lavoura, M.~N.~Rebelo, M.~Sher and J.~P.~Silva, ``Theory and phenomenology of two-Higgs-doublet models'', \href{https://www.sciencedirect.com/science/article/abs/pii/S0370157312000695?via/3Dihub}{\underline{Phys. Rept.} \textbf{516} (2012) 1-102}, \href{https://arxiv.org/pdf/1106.0034.pdf}{\texttt{arXiv:1106.0034 [hep-ph]}}.\vspace{0.1cm}
	
	\bibitem{CMS:2018cyk}
	CMS Collaboration, A.~M.~Sirunyan \underline{et al.}, ``Search for a standard model-like Higgs boson in the mass range between 70 and 110 GeV in the diphoton final state in proton-proton collisions at $\sqrt{s}=$ 8 and 13 TeV'' ,\href{https://www.sciencedirect.com/science/article/pii/S0370269319302904}{\underline{Phys. Lett.} \textbf{B793} (2019) 320-347}, \href{https://arxiv.org/pdf/1811.08459.pdf}{\texttt{arXiv:1811.08459 [hep-ex]}}.\vspace{0.1cm}
	
	\bibitem{CMS:2022rbd}
	CMS Collaboration, ``Searches for additional Higgs bosons and vector-like leptoquarks in $\tau \tau$ final states in proton-proton collisions at $\sqrt{s} $ = 13TeV'',\href{https://inspirehep.net/files/8922ea5ea5359a7942834b9b1f0a577f}{CMS-PAS-HIG-21-001}.\vspace{0.1cm}
	
	\bibitem{Cao:2016uwt}
	J.~Cao, X.~Guo, Y.~He, P.~Wu and Y.~Zhang, ``Diphoton signal of the light Higgs boson in natural NMSSM'', \href{https://journals.aps.org/prd/abstract/10.1103/PhysRevD.95.116001}{\underline{Phys. Rev.}  \textbf{D95} (2017) no.11, 116001}, \href{https://arxiv.org/pdf/1612.08522.pdf}{arXiv:1612.08522 [hep-ph]}.\vspace{0.1cm}
	
	\bibitem{Heinemeyer:2021msz}
	S.~Heinemeyer, C.~Li, F.~Lika, G.~Moortgat-Pick and S.~Paasch,``A 96 GeV Higgs Boson in the 2HDM plus Singlet''
	\href{https://arxiv.org/pdf/2112.11958.pdf}{\texttt{arXiv:2112.11958 [hep-ph]}}.\vspace{0.1cm}
	
	\bibitem{Biekotter:2021qbc}
	T.~Biek\"otter, A.~Grohsjean, S.~Heinemeyer, C.~Schwanenberger and G.~Weiglein, ``Possible indications for new Higgs bosons in the reach of the LHC: N2HDM and NMSSM interpretations'', \href{https://arxiv.org/pdf/2109.01128.pdf}{\texttt{arXiv:2109.01128 [hep-ph]}}.\vspace{0.1cm}
	
	
	\bibitem{Biekotter:2019kde}
	T.~Biek\"otter, M.~Chakraborti and S.~Heinemeyer, ``A 96 GeV Higgs boson in the N2HDM'', \href{https://link.springer.com/article/10.1140/2Fepjc/2Fs10052-019-7561-2}{\underline{Eur. Phys. J.}  \textbf{C80} (2020) no.1, 2}, \href{https://arxiv.org/pdf/1903.11661.pdf}{\texttt{arXiv:1903.11661 [hep-ph]}}.\vspace{0.1cm}
	
	\bibitem{Cao:2019ofo}
	J.~Cao, X.~Jia, Y.~Yue, H.~Zhou and P.~Zhu, ``96 GeV diphoton excess in seesaw extensions of the natural NMSSM'', \href{https://journals.aps.org/prd/abstract/10.1103/PhysRevD.101.055008}{\underline{Phys. Rev.}  \textbf{D101} (2020) no.5, 055008}, \href{https://arxiv.org/pdf/1908.07206.pdf}{\texttt{arXiv:1908.07206 [hep-ph]}}.\vspace{0.1cm}
	
	\bibitem{Biekotter:2022jyr}
	T.~Biek\"otter, S.~Heinemeyer and G.~Weiglein, ``Mounting evidence for a 95 GeV Higgs boson'', \href{https://arxiv.org/pdf/2203.13180.pdf}{arXiv:2203.13180 [hep-ph]}.\vspace{0.1cm}
	
	\bibitem{Biekotter:2022abc}
	T.~Biek\"otter, S.~Heinemeyer and G.~Weiglein, ``Excesses in the low-mass Higgs-boson search and the W-boson mass measurement'', \href{https://arxiv.org/pdf/2204.05975.pdf}{\texttt{arXiv:2204.05975 [hep-ph]}}.\vspace{0.1cm}
	
	
	\bibitem{Cline:2019okt}
	J.~M.~Cline and T.~Toma, ``Pseudo-Goldstone dark matter confronts cosmic ray and collider anomalies''
	, \href{https://journals.aps.org/prd/abstract/10.1103/PhysRevD.100.035023}{\underline{Phys. Rev.}  \textbf{D100} (2019) no.3, 035023} \href{https://arxiv.org/pdf/1906.02175.pdf}{\texttt{arXiv:1906.02175 [hep-ph]}}.\vspace{0.1cm}
	
	
	\bibitem{Biekotter:2021ovi}
	T.~Biek\"otter and M.~O.~Olea-Romacho, ``Reconciling Higgs physics and pseudo-Nambu-Goldstone dark matter in the S2HDM using a genetic algorithm'', \href{https://link.springer.com/article/10.1007/JHEP10(2021)215}{\underline{JHEP} \textbf{10} (2021), 215}, \href{https://arxiv.org/pdf/2108.10864.pdf}{\texttt{arXiv:2108.10864 [hep-ph]}}.\vspace{0.1cm}
	
	\bibitem{Crivellin:2017upt}
	A.~Crivellin, J.~Heeck and D.~M\"uller, ``Large $h\to b s$ in generic two-Higgs-doublet models'', \href{https://journals.aps.org/prd/abstract/10.1103/PhysRevD.97.035008}{\underline{Phys. Rev.}  \textbf{D97} (2018) no.3, 035008}, \href{https://arxiv.org/pdf/1710.04663.pdf}{\texttt{arXiv:1710.04663 [hep-ph]}}.\vspace{0.1cm}
	
	
	\bibitem{ALEPH:2006tnd}
	S.~Schael \textit{et al.} [ALEPH, DELPHI, L3, OPAL and LEP Working Group for Higgs Boson Searches], ``Search for neutral MSSM Higgs bosons at LEP'', \href{https://link.springer.com/article/10.1140/2Fepjc/2Fs2006-02569-7}{\underline{Eur. Phys. J.}  \textbf{C47} (2006), 547-587}, \href{https://arxiv.org/pdf/hep-ex/0602042.pdf}{\texttt{arXiv:hep-ex/0602042 [hep-ex]}}.\vspace{0.1cm}
	
	
	\bibitem{Cacciapaglia:2016tlr}
	G.~Cacciapaglia, A.~Deandrea, S.~Gascon-Shotkin, S.~Le Corre, M.~Lethuillier and J.~Tao, ``Search for a lighter Higgs boson in Two Higgs Doublet Models'', \href{https://dx.doi.org/10.1007/JHEP12(2016)068}{\underline{JHEP} \textbf{12} (2016) 068}, \href{https://arxiv.org/pdf/1607.08653.pdf}{\texttt{arXiv:1607.08653 [hep-ph]}}.\vspace{0.1cm}
	
	\bibitem{Abdelalim:2020xfk}
	A.~A.~Abdelalim, B.~Das, S.~Khalil and S.~Moretti, ``Di-photon decay of a light Higgs state in the BLSSM'' \href{https://arxiv.org/pdf/2012.04952.pdf}{\texttt{arXiv:2012.04952 [hep-ph]}}.\vspace{0.1cm}
	
	\bibitem{LEP-Excess}
	R.~Barate \textit{et al.} [LEP Working Group for Higgs boson searches, ALEPH, DELPHI, L3 and OPAL], ``Search for the standard model Higgs boson at LEP'', \href{https://www.sciencedirect.com/science/article/pii/S0370269303006142?via\%3Dihub}{\underline{Phys. Lett.}  \textbf{B565} (2003), 61-75}, \href{https://arxiv.org/pdf/hep-ex/0306033.pdf}{\texttt{arXiv:hep-ex/0306033 [hep-ex]}}.\vspace{0.1cm}
	
	\bibitem{Crivellin:2013wna}
	A.~Crivellin, A.~Kokulu and C.~Greub,``Flavor-phenomenology of two-Higgs-doublet models with generic Yukawa structure''
	\href{https://journals.aps.org/prd/abstract/10.1103/PhysRevD.87.094031}{\underline{Phys. Rev.}  \textbf{D87} (2013) no.9, 094031}, \href{Uhttps://arxiv.org/pdf/1303.5877.pdfRL}{\texttt{arXiv:1303.5877 [hep-ph]}}.\vspace{0.1cm}
	
	
	
	
	\bibitem{Cheng:1987rs}
	T.~P.~Cheng and M.~Sher, ``Mass Matrix Ansatz and Flavor Nonconservation in Models with Multiple Higgs Doublets'', \href{https://journals.aps.org/prd/abstract/10.1103/PhysRevD.35.3484}{\underline{Phys. Rev.} \textbf{D35} (1987) 3484}.\vspace{0.1cm}
	
	\bibitem{Diaz-Cruz:2004wsi}
	J.~L.~Diaz-Cruz, R.~Noriega-Papaqui and A.~Rosado, ``Mass matrix ansatz and lepton flavour violation in the 2HDM-III'', \href{https://journals.aps.org/prd/abstract/10.1103/PhysRevD.69.095002}{\underline{Phys. Rev.}  \textbf{D69} (2004), 095002}\href{https://arxiv.org/pdf/hep-ph/0401194v2.pdf}{\texttt{arXiv:hep-ph/0401194 [hep-ph]}}.\vspace{0.1cm}
	
	\bibitem{Benbrik:2015evd}  
	R.~Benbrik, C.~H.~Chen and T.~Nomura, ``$h,Z\to \ell_i \bar\ell_j$, $\Delta a_{\mu}$, $\tau\to (3\mu,\mu \gamma)$ in generic two-Higgs-doublet models'', \href{https://journals.aps.org/prd/abstract/10.1103/PhysRevD.63.091301}{\underline{Phys. Rev.}  \textbf{D93} (2016) no.9, 095004}, \href{https://arxiv.org/pdf/hep-ph/0009292.pdf}{\texttt{arXiv:hep-ph/0009292 [hep-ph]}}.\vspace{0.1cm}
	
	
	\bibitem{Xie:2018yiv} 
	W.~Xie, R.~Benbrik, A.~Habjia, S.~Taj, B.~Gong and Q.~S.~Yan, ``Signature of 2HDM at Higgs Factories'', \href{https://journals.aps.org/prd/abstract/10.1103/PhysRevD.103.095030}{\underline{Phys. Rev.}  \textbf{D103} (2021) no.9, 095030}, \href{https://arxiv.org/pdf/1812.02597.pdf}{\texttt{arXiv:1812.02597 [hep-ph]}}.\vspace{0.1cm}	
	
	\bibitem{Kanemura:2017gbi}
	S.~Kanemura, M.~Kikuchi, K.~Sakurai and K.~Yagyu, ``H-COUP: a program for one-loop corrected Higgs boson couplings in non-minimal Higgs sectors'', \href{https://www.sciencedirect.com/science/article/abs/pii/S0010465518302200?via%3Dihub}{\underline{Comput. Phys. Commun.} \textbf{233} (2018), 134-144}, \href{https://arxiv.org/pdf/1710.04603.pdf}{\texttt{arXiv:1710.04603 [hep-ph]}}.\vspace{0.1cm}
	
	
	
	
	
	
	
	\bibitem{2HDMC}
	D.~Eriksson, J.~Rathsman and O.~Stal, ``2HDMC: Two-Higgs-Doublet Model Calculator Physics and Manual,''
	\href{https://www.sciencedirect.com/science/article/abs/pii/S0010465509003014?via/3Dihub}{\underline{Comput. Phys. Commun.} \textbf{181} (2010) 189-205}, \href{https://arxiv.org/pdf/0902.0851.pdf}{\texttt{arXiv:0902.0851 [hep-ph]}}.\vspace{0.1cm}
	
	
	
	\bibitem{uni1}
	S.~Kanemura, T.~Kubota and E.~Takasugi, ``Lee-Quigg-Thacker bounds for Higgs boson masses in a two doublet model'', \href{https://www.sciencedirect.com/science/article/abs/pii/0370269393912052?via/3Dihub}{\underline{Phys. Lett.} \textbf{B313} (1993) 155-160}, \href{https://arxiv.org/pdf/hep-ph/9303263.pdf}{\texttt{arXiv:hep-ph/9303263 [hep-ph]}}.\vspace{0.3cm}
	
	\bibitem{uni2}
	A.~G.~Akeroyd, A.~Arhrib and E.~M.~Naimi, ``Note on tree level unitarity in the general two Higgs doublet model'',  \href{https://www.sciencedirect.com/science/article/abs/pii/S037026930000962X?via/3Dihub}{\underline{Phys. Lett.} \textbf{B490} (2000) 119-124}, \href{https://arxiv.org/pdf/hep-ph/0006035.pdf}{\texttt{arXiv:hep-ph/0006035 [hep-ph]}}.\vspace{0.3cm}
	
	\bibitem{uni3}
	A.~Arhrib, ``Unitarity constraints on scalar parameters of the standard and two Higgs doublets model'', \href{https://arxiv.org/pdf/hep-ph/0012353.pdf}{\texttt{arXiv:hep-ph/0012353 [hep-ph]}}.\vspace{0.3cm}
	
	\bibitem{Barroso:2013awa}
	A.~Barroso, P.~M.~Ferreira, I.~P.~Ivanov and R.~Santos, ``Metastability bounds on the two Higgs doublet model'', \href{https://doi.org/10.1007/JHEP06(2013)045}{\underline{JHEP} \textbf{06} (2013), 045}, \href{https://arxiv.org/pdf/1303.5098.pdf}{\texttt{arXiv:1303.5098 [hep-ph]}}.\vspace{0.1cm}
	
	\bibitem{sta}
	N.~G.~Deshpande and E.~Ma, ``Pattern of Symmetry Breaking with Two Higgs Doublets'', \href{https://journals.aps.org/prd/abstract/10.1103/PhysRevD.18.2574}{\underline{Phys. Rev.} \textbf{D18} (1978) 2574}.\vspace{0.1cm}
	
	\bibitem{Grimus:2007if}
	W.~Grimus, L.~Lavoura, O.~M.~Ogreid and P.~Osland, ``A Precision constraint on multi-Higgs-doublet models'', \href{https://iopscience.iop.org/article/10.1088/0954-3899/35/7/075001}{\underline{J. Phys.}  \textbf{G35} (2008), 075001}, \href{https://arxiv.org/pdf/0711.4022.pdf}{\texttt{arXiv:0711.4022 [hep-ph]}}.\vspace{0.1cm}
	
	
	\bibitem{oblique2}
	W.~Grimus, L.~Lavoura, O.~M.~Ogreid and P.~Osland, ``The Oblique parameters in multi-Higgs-doublet models'', \href{https://www.sciencedirect.com/science/article/abs/pii/S0550321308002289?via/3Dihub}{\underline{Nucl. Phys.}  \textbf{B801} (2008) 81-96}, \href{https://arxiv.org/pdf/0802.4353.pdf}{\texttt{arXiv:0802.4353 [hep-ph]}}.\vspace{0.1cm}
	
	\bibitem{particle2020review} 
	{Particle Data Group} Collaboration, P.~Zyla {et al.}, ``Review of Particle Physics'', \href{https://academic.oup.com/ptep/article/2020/8/083C01/5891211}{\underline{Progress of Theoretical and Experimental Physics} \textbf{2020}  no. 8 (2020) 083C01}.\vspace{0.1cm}
	
	
	\bibitem{Bechtle:2008jh}
	P.~Bechtle, O.~Brein, S.~Heinemeyer, G.~Weiglein and K.~E.~Williams, ``HiggsBounds: Confronting Arbitrary Higgs Sectors with Exclusion Bounds from LEP and the Tevatron'', \href{https://linkinghub.elsevier.com/retrieve/pii/S0010465509002823}{\underline{Comput. Phys. Commun.} \textbf{181} (2010), \href{https://arxiv.org/pdf/0811.4169.pdf}{\texttt{arXiv:0811.4169 [hep-ph]} 138-167}}.\vspace{0.1cm}
	
	\bibitem{Bechtle:2011sb}
	P.~Bechtle, O.~Brein, S.~Heinemeyer, G.~Weiglein and K.~E.~Williams, ``HiggsBounds 2.0.0: Confronting Neutral and Charged Higgs Sector Predictions with Exclusion Bounds from LEP and the Tevatron'', \href{https://linkinghub.elsevier.com/retrieve/pii/S0010465511002621}{underline{Comput. Phys. Commun.} \textbf{182} (2011), 2605-2631}, \href{https://arxiv.org/pdf/1102.1898.pdf}{\texttt{arXiv:1102.1898 [hep-ph]}}.\vspace{0.1cm}
	
	\bibitem{Bechtle:2013wla}
	P.~Bechtle, O.~Brein, S.~Heinemeyer, O.~St\r{a}l, T.~Stefaniak, G.~Weiglein and K.~E.~Williams,``$\mathsf{HiggsBounds}-4$: Improved Tests of Extended Higgs Sectors against Exclusion Bounds from LEP, the Tevatron and the LHC'', \href{https://link.springer.com/article/10.1140/2Fepjc/2Fs10052-013-2693-2}{\underline{Eur. Phys. J.}  \textbf{C74} (2014) no.3, 2693}, \href{https://arxiv.org/pdf/1311.0055.pdf}{\texttt{arXiv:1311.0055 [hep-ph]}}.\vspace{0.1cm}
	
	\bibitem{Bechtle:2014ewa}
	P.~Bechtle, S.~Heinemeyer, O.~St\r{a}l, T.~Stefaniak and G.~Weiglein, ``Probing the Standard Model with Higgs signal rates from the Tevatron, the LHC and a future ILC'', \href{https://link.springer.com/article/10.1007/2FJHEP11/282014/29039}{\underline{JHEP} \textbf{11} (2014), 039}, \href{https://arxiv.org/pdf/1403.1582.pdf}{\texttt{arXiv:1403.1582 [hep-ph]}}.\vspace{0.1cm}
	
	\bibitem{Bechtle:2015pma}
	P.~Bechtle, S.~Heinemeyer, O.~Stal, T.~Stefaniak and G.~Weiglein, ``Applying Exclusion Likelihoods from LHC Searches to Extended Higgs Sectors'', \href{https://doi.org/10.1140/epjc/s10052-015-3650-z}{\underline{Eur. Phys. J}.  \textbf{C75} (2015) no.9, 421}, \href{https://arxiv.org/pdf/1507.06706}{\texttt{arXiv:1507.06706 [hep-ph]}}.\vspace{0.1cm}
	
	\bibitem{Bechtle:2020pkv}
	P.~Bechtle, D.~Dercks, S.~Heinemeyer, T.~Klingl, T.~Stefaniak, G.~Weiglein and J.~Wittbrodt, ``HiggsBounds-5: Testing Higgs Sectors in the LHC 13 TeV Era'', \href{https://link.springer.com/article/10.1140/2Fepjc/2Fs10052-020-08557-9}{\underline{Eur. Phys. J.}  \textbf{C80} (2020) 1211}, \href{https://arxiv.org/pdf/2006.06007.pdf}{\texttt{arXiv:2006.06007 [hep-ph]}}.\vspace{0.1cm}
	
	
	\bibitem{HS1}
	P.~Bechtle, S.~Heinemeyer, O.~St\r{a}l, T.~Stefaniak and G.~Weiglein,``$HiggsSignals$: Confronting arbitrary Higgs sectors with measurements at the Tevatron and the LHC'', \href{https://link.springer.com/article/10.1140/2Fepjc/2Fs10052-013-2711-4}{\underline{Eur. Phys. J.} \textbf{C74} (2014) no.2, 2711}, \href{https://arxiv.org/pdf/1305.1933.pdf}{\texttt{arXiv:1305.1933 [hep-ph]}}.\vspace{0.1cm}
	
	
	\bibitem{HS}
	P.~Bechtle, S.~Heinemeyer, T.~Klingl, T.~Stefaniak, G.~Weiglein and J.~Wittbrodt, ``HiggsSignals-2: Probing new physics with precision Higgs measurements in the LHC 13 TeV era'', \href{https://link.springer.com/article/10.1140/2Fepjc/2Fs10052-021-08942-y}{\underline{Eur. Phys. J.}  \textbf{C81} (2021) 145},  \href{https://arxiv.org/pdf/2012.09197.pdf}{\texttt{arXiv:2012.09197 [hep-ph]}}.\vspace{0.1cm}
	
	
	
	\bibitem{superIso}
	F.~Mahmoudi, ``SuperIso v2.3: A Program for calculating flavor physics observables in Supersymmetry'', \href{https://www.sciencedirect.com/science/article/abs/pii/S0010465509000721?via/3Dihub}{\underline{Comput. Phys. Commun.} \textbf{180} (2009) 1579-1613}, \href{https://arxiv.org/pdf/0808.3144.pdf}{\texttt{arXiv:0808.3144 [hep-ph]}}.\vspace{0.1cm}
	
	
	\bibitem{Amhis:2014hma}
	{Heavy Flavor Averaging Group (HFAG)}, Y.~Amhis \underline{et al.}, ``Averages of $b$-hadron, $c$-hadron, and $\tau$-lepton properties as of summer 2014'', \href{https://arxiv.org/pdf/1412.7515.pdf}{\texttt{arXiv:1412.7515 [hep-ex]}}.\vspace{0.3cm}
	
	\bibitem{Bphy1}
	{Heavy Flavor Averaging Group}, \url{https://www.slac.stanford.edu/xorg/hfag/rare/2013/radll/btosg.pdf}.\vspace{0.1cm}
	
	\bibitem{Bphy2}
	{Heavy Flavor Averaging Group}, \url{https://www.slac.stanford.edu/xorg/hfag/rare/2014/bs/OUTPUT/TABLES/bs.pdf}.\vspace{0.1cm}
	
	
	\bibitem{Archilli:2014cla}
	F.~Archilli, ``$B^0_{s} \rightarrow \mu^+\mu^-$ at LHC'', \href{https://arxiv.org/pdf/1411.4964.pdf}{\texttt{arXiv:1411.4964 [hep-ex]}}.\vspace{0.1cm}
	
	
	\bibitem{Bphy3}
	{Heavy Flavor Averaging Group}, \url{https://www.slac.stanford.edu/xorg/hfag/rare/2014/radll/OUTPUT/HTML/radll_table4.html}.\vspace{0.1cm}
	
	\bibitem{ParticleDataGroup:2008zun}
	{Particle Data Group} Collaboration, C.~Amsler \textit{et al.},``Review of Particle Physics''
	\href{https://linkinghub.elsevier.com/retrieve/pii/S0370269308008435}{\underline{Phys. Lett.}  \textbf{B667} (2008), 1-1340}.\vspace{0.1cm}
	
	\bibitem{FlaviaNetWorkingGrouponKaonDecays:2008hpm}
	{FlaviaNet Working Group on Kaon Decays}, M.~Antonelli \textit{et al.}, ``Precision tests of the Standard Model with leptonic and semileptonic kaon decays'', \href{https://arxiv.org/pdf/0801.1817.pdf}{\texttt{arXiv:0801.1817 [hep-ph]}}.\vspace{0.1cm}
	
	\bibitem{Miller:2007kk}
	J.~P.~Miller, E.~de Rafael and B.~L.~Roberts, ``Muon (g-2): Experiment and theory''
	\href{https://iopscience.iop.org/article/10.1088/0034-4885/70/5/R03}{\underline{Rept. Prog. Phys.} \textbf{70} (2007), 795}, \href{https://arxiv.org/pdf/hep-ph/0703049.pdf}{\texttt{arXiv:hep-ph/0703049 [hep-ph]}}.\vspace{0.1cm}
	
	\bibitem{Chang:2000ii}
	D.~Chang, W.~F.~Chang, C.~H.~Chou and W.~Y.~Keung, ``Large two loop contributions to g-2 from a generic pseudoscalar boson'', \href{https://journals.aps.org/prd/abstract/10.1103/PhysRevD.63.091301}{\underline{Phys. Rev.}  \textbf{D63} (2001), 091301}, \href{https://arxiv.org/pdf/hep-ph/0009292.pdf}{\texttt{[arXiv:hep-ph/0009292 [hep-ph]]}}.\vspace{0.1cm}



\bibitem{Cheung:2001hz}
K.~m.~Cheung, C.~H.~Chou and O.~C.~W.~Kong, ``Muon anomalous magnetic moment, two Higgs doublet model, and supersymmetry'', \href{https://journals.aps.org/prd/abstract/10.1103/PhysRevD.64.111301}{\underline{Phys. Rev.}  \textbf{D64} (2001), 111301ext}, \href{https://arxiv.org/pdf/hep-ph/0103183.pdf}{\texttt{arXiv:hep-ph/0103183 [hep-ph]}}.\vspace{0.1cm}


\bibitem{Misiak:2017bgg}
M.~Misiak and M.~Steinhauser,``Weak radiative decays of the B meson and bounds on $M_{H^\pm }$ in the Two-Higgs-Doublet Model''
\href{https://link.springer.com/article/10.1140/epjc/s10052-017-4776-y}{\underline{Eur. Phys. J.}  \textbf{C77} (2017) no.3, 201}, \href{https://arxiv.org/pdf/1702.04571.pdf}{\texttt{arXiv:1702.04571 [hep-ph]}}.\vspace{0.1cm}




\bibitem{CDF:2013jor}
T.~Aaltonen \textit{et al.} [CDF],``Updated search for the standard model Higgs boson in events with jets and missing transverse energy using the full CDF data set'', \href{https://journals.aps.org/prd/abstract/10.1103/PhysRevD.87.052008}{\underline{Phys. Rev.}  \textbf{D87} (2013), 052008}, \href{https://arxiv.org/pdf/1301.4440.pdf}{\texttt{arXiv:1301.4440 [hep-ex]}}.\vspace{0.1cm}

\bibitem{ATLAS:2018xad}
ATLAS Collaboration, ``Search for resonances in the 65 to 110 GeV diphoton invariant mass range using 80 fb$^{-1}$ of $pp$ collisions collected at $\sqrt{s}=13$ TeV with the ATLAS detector'',\href{http://cdsweb.cern.ch/record/2628760/files/ATLAS-CONF-2018-025.pdf}{\texttt{ATLAS-CONF-2018-025}}
.\vspace{0.1cm}



\end{thebibliography}
\end{document}